\documentclass{amsart}
\usepackage[utf8]{inputenc}
\usepackage{amsfonts}
\usepackage{amsmath}
\usepackage{natbib}
\usepackage{bm}
\usepackage[usenames,dvipsnames,svgnames,table]{xcolor}
\usepackage[font=footnotesize,labelfont=bf]{caption}
\usepackage{colortbl}
\usepackage{booktabs}
\usepackage{fixltx2e}
\usepackage{graphicx}
\usepackage{float}
\usepackage{bbm}
\usepackage[hidelinks]{hyperref}
\usepackage{standalone}
\usepackage{dubious}
\usepackage{ulem}

\usepackage{indentfirst}

\RequirePackage{calc}

\newcommand{\beginsupplement}{%
        \setcounter{table}{0}
        \renewcommand{\thetable}{S\arabic{table}}%
        \setcounter{figure}{0}
        \renewcommand{\thefigure}{S\arabic{figure}}%
     }

\definecolor{lightergray}{rgb}{0.9, 0.9, 0.9}

\newcommand{\notarxiv}[1]{}
\newcommand{\arxiv}[1]{#1}

\begin{document}

\arxiv{
\bigskip
\noindent RH: THE MARGINAL LIKELIHOOD OF A TREE TOPOLOGY
\bigskip
\medskip

\title[The marginal likelihood of a tree topology]{19 dubious ways to compute the marginal likelihood of a phylogenetic tree topology}

\begin{center}

\author[M. Fourment, A. Magee, C. Whidden, A. Bilge, F. A. Matsen~IV, V. N. Minin]{Mathieu Fourment$^1$, Andrew F. Magee$^{2}$, Chris Whidden$^{3}$, Arman Bilge$^3$, \\ Frederick A. Matsen~IV$^{3,\ast}$, Vladimir N. Minin$^{4,\ast}$}

\maketitle

\noindent {\small \it
 $^1$University of Technology Sydney, ithree institute, Ultimo NSW 2007, Australia\\
 $^2$Department of Biology, University of Washington, Seattle, WA, 98195, USA\\
 $^3$Fred Hutchinson Cancer Research Center, Seattle, WA, 98109, USA\\
 $^4$Department of Statistics, University of California, Irvine, CA, 92697, USA}
 \end{center}

\medskip
 \noindent{\bf Corresponding authors:} Frederick A. Matsen~IV, Fred Hutchinson Cancer Research Center, 1100 Fairview Ave.\ N, Mail stop M1-B514, Seattle, WA, 98109, USA; E-mail: matsen@fredhutch.org \& Vladimir Minin, Department of Statistics, University of California, Irvine, CA, 92697, USA; E-mail: vminin@uci.edu\\
}

\begin{abstract}
{The marginal likelihood of a model is a key quantity for assessing the evidence provided by the data in support of a model.
The marginal likelihood is the normalizing constant for the posterior density, obtained by integrating the product of the likelihood and the prior with respect to model parameters.
Thus, the computational burden of computing the marginal likelihood scales with the dimension of the parameter space.
In phylogenetics, where we work with tree topologies that are high-dimensional models, standard approaches to computing marginal likelihoods are very slow.
Here we study methods to quickly compute the marginal likelihood of a single fixed tree topology.
We benchmark the speed and accuracy of 19 different methods to compute the marginal likelihood of phylogenetic topologies on a suite of real datasets.
These methods include several new ones that we develop explicitly to solve this problem, as well as existing algorithms that we apply to phylogenetic models for the first time.
Altogether, our results show that the accuracy of these methods varies widely, and that accuracy does not necessarily correlate with computational burden.
Our newly developed methods are orders of magnitude faster than standard approaches, and in some cases, their accuracy rivals the best established estimators.}
\notarxiv{{Bayesian inference, model selection, evidence}}
\end{abstract}
%\newline

\arxiv{
\newline

\section*{Introduction}
}
In phylogenetic inference, the tree topology forms a key object of inference.
In Bayesian phylogenetics, this translates to approximating the posterior distribution of tree topologies.
Typically, a joint posterior distribution of tree topologies and continuous parameters, including branch lengths and substitution model parameters, is approximated directly via Markov chain Monte Carlo (MCMC), as done in the popular Bayesian phylogenetics software MrBayes \citep{ronquist2012mrbayes}.
However, MCMC over topologies is computationally expensive \citep{lakner2008efficiency,hohna2008clock}.
These MCMC algorithms spend a nontrivial amount of time marginalizing over branch lengths and substitution models parameters and discarding them so that the estimated posterior probability of a tree topology is the proportion of MCMC iterations in which it appears.
Therefore, fast marginalization over continuous phylogenetic parameters may offer a boon to MCMC algorithm efficiency or even allow one to perform Bayesian phylogenetic inference without MCMC.
In this paper, we  review existing methods and develop new ones to compute the posterior probabilities of tree topologies by quickly marginalizing out branch lengths to compute the marginal likelihood of a given topology.
We compare speed and accuracy of 19 methods and examine whether there is a speed-accuracy trade off.

Given that the bulk of Bayesian inference is performed with methods that work because they allow the marginal likelihood to be avoided, why would one want to compute them at all?
One potential application of these marginal likelihood computations is the development of fast, MCMC-free Bayesian phylogenetic inference.
To make such an advance, first one would need to identify a large enough set of \textit{a posteori} highly probable tree topologies, such as with a new optimization-based method called phylogenetic topographer (PT) \citep{pt}.
Once a set of promising tree topologies is formed, we can compute their marginal likelihoods, then renormalize these marginal likelihoods (perhaps after multiplying by a prior) to obtain approximate posterior probabilities of tree topologies --- the key output of Bayesian phylogenetic inference.
Luckily, we can tap into a substantial body of research on computing the marginal likelihood of purely continuous statistical models in order to integrate out continuous parameters for any given tree topology \citep{hans2007shotgun,lenkoski2011computational}.
It is therefore high time we consider the possibility of constructing the posterior distribution on topologies without MCMC.
To do so, we must know: how well, and how quickly, can we compute the marginal likelihood of a topology?

In this paper, we address this question by benchmarking a wide range of methods for calculating the marginal likelihood of a topology with respect to branch lengths under the JC69 model, the simplest nucleotide substitution model.
These approaches include very fast approximations including several based on the Laplace approximation \citep{tierney1986accurate,kass1995bayes} and variational approaches \citep{ranganath2014black}.
There are also approaches that require some sampling (though not of topologies), including those that make use of MCMC samples (\CF\ bridge sampling, \citep{overstall2010default,gronau2017tutorial}) and approaches that employ importance sampling (\CF\ na{\"i}ve Monte Carlo, \citep{hammersley1964general,raftery1991stopping}).
We also include approaches that make use of a set of so-called power posteriors, including the path sampling \citep{ogata1989monte,gelman1998simulating,lartillot2006computing,baele2012improving} method frequently used in phylogenetics.
Using a set of empirical datasets and a common inference framework, we benchmark 19 methods for computing the marginal likelihood of tree topologies.
These 19 methods include some well-known in the phylogenetics literature, some we apply for the first time in phylogenetics, and others that we develop explicitly for this problem.
We find that some of these new methods provide estimates that compare favorably to the precise (but slow) state-of-the-art approaches, while running orders of magnitudes more quickly.
The title of our paper is adapted from the classic review of matrix exponentiation methods by \citet{Moler1978-gi, Moler2003-au}; it is not meant to cast doubt on the methods presented here, although we do find that some rather ``dubious'' methods making strong simplifying assumptions perform surprisingly well!
\bigskip

\section*{Methods}

\begin{table}[H]
\small
\caption{Names, abbreviations, and number of required MCMC chains involved in applying the 19 methods. \GLIS, \VBIS, and NMC ($^*$) do not require MCMC samples but perform importance sampling. Stepping stone and path sampling methods employ an un-specified number of steps; we found 50 to be sufficient.}
\label{tab:method_names}
\centering
\begin{tabular} {lll}
\toprule
Abbreviation 		& Full name 	& $\#$ MCMC chains\\
\midrule
\rowcolor{lightergray} ELBO & Evidence Lower Bound & 0 \\
\rowcolor{white} \GLIS & Gamma Laplus Importance Sampling & 0$^*$ \\
\rowcolor{lightergray} \VBIS & Varational Bayes Importance Sampling  & 0$^*$ \\
\rowcolor{white} BL & \Betaprime\ Laplus & 0 \\
\rowcolor{lightergray} \GL & Gamma Laplus & 0 \\
\rowcolor{white} \LL & Lognormal Laplus & 0 \\
\rowcolor{lightergray} MAP & Maximum un-normalized posterior probability & 0 \\
\rowcolor{white} ML & Maximum likelihood & 0 \\
\rowcolor{lightergray} NMC & Na\"{i}ve Monte Carlo & 0$^*$ \\
\rowcolor{white} BS & Bridge Sampling & 1\\
\rowcolor{lightergray} CPO & Conditional Predictive Ordinates & 1\\
\rowcolor{white} HM & Harmonic Mean & 1 \\
\rowcolor{lightergray} \SHM & Stabilized Harmonic Mean & 1 \\
\rowcolor{white} NS & Nested Sampling & multiple short chains \\
\rowcolor{lightergray} \LPPD & Pointwise Predictive Density & 1 \\
\rowcolor{white} \PS & Path Sampling & 50 \\
\rowcolor{lightergray} \PStwo & Modified Path Sampling & 50 \\
\rowcolor{white} \SStone & Stepping Stone & 50 \\
\rowcolor{lightergray} \GSS & Generalized Stepping Stone & 50 \\
\bottomrule
\end{tabular}
\end{table}

\subsection*{Marginal likelihoods}
Consider a (fixed) unrooted topology $\tau$ for $S$ species with unconstrained branch length vector $\boldsymbol{\theta} = (\theta_1, \theta_2, \dots ,\theta_{2S-3})$ and the JC69 (Jukes-Cantor) model \citep{jukes1969evolution}.
The JC69 model does not any have free parameters as it assumes equal base frequencies and equal substitution rate for all pairs of nucleotides.
If branch lengths are measured in units of the expected number of substitutions per site and the JC69 substitution model is employed, the posterior distribution is given by:
\[
p(\boldsymbol{\theta} \mid \tau, D) = \frac{p(D \mid \boldsymbol{\theta}, \tau) p(\boldsymbol{\theta}\mid \tau)}{\int_{[0,\infty]^{2S-3}} p(D \mid \boldsymbol{\theta}, \tau) p(\boldsymbol{\theta} \mid \tau) \text{d}\boldsymbol{\theta}}.
\]
The normalizing constant in the denominator of the right hand side is the marginal likelihood of the phylogenetic tree topology model $\tau$, $p(D \mid \tau)$.
It is this marginal likelihood (of a sequence alignment given a topology) that is the quantity of interest in this manuscript.
As is typical, we place independent exponential priors on branch lengths with a prior expectation of 0.1 substitutions, such that $p(\boldsymbol{\theta} \mid \tau) = p(\boldsymbol{\theta}) = \prod_{i=1}^{2S-3} p(\theta_i)$, where $p(x)$ is the exponential density.

Calculating marginal likelihoods is an area of active statistical research, both inside and outside of phylogenetics.
A complete review of all the methods that have been proposed for this purpose is outside the scope of this paper, and we refer readers to reviews by \citet{gelman1998simulating} and \citet{gronau2017tutorial}.
We will first provide a basic sketch of the types of methods we employ (see Table \ref{tab:method_names} for abbreviations).
Second, we describe some new methods for calculating the marginal likelihood designed specifically for topologies.
Finally, a more detailed explanation of all the methods used in this paper can be found in the supplementary materials.

Methods for calculating the marginal likelihood can be broken down into two main categories: sampling-free methods and sampling-based methods.
The majority of sampling-free methods revolve around replacing the intractable posterior distribution with one whose normalizing constant can be more easily computed.
These approaches include the Laplace approximation \citep{tierney1986accurate,kass1995bayes}, three new variations on this theme that we introduce here (the Laplus approximations), and a variational Bayes approximation \citep{ranganath2014black} from which we derive the evidence lower bound (ELBO).
We additionally investigate the performance of the maximum likelihood and maximum a posteriori estimators to approximate the marginal likelihood.

The sampling-based approaches can further be broken down into importance sampling and MCMC-based aproaches.
In importance sampling, samples drawn from a tractable proposal distribution are used to calculate the marginal likelihood using simple identities.
How well an importance sampling method works depends on how close the proposal distribution is to the true posterior.
We examine three importance sampling approaches, na{\"i}ve Monte Carlo (NMC) \citep{hammersley1964general,raftery1991stopping}, which uses the prior distribution as the proposal distribution, and two approaches using more sophisticated proposal distributions.
Lastly, the MCMC-based methods can be broken down into those that can be used with a single chain, and those that require many chains.
Among single-chain methods, we include the well-known harmonic mean (HM) estimator \citep{newton1994approximate}, a variation thereof known as the stabilized harmonic mean (\SHM) \citep{newton1994approximate}, bridge sampling (BS) \citep{overstall2010default,gronau2017tutorial}, conditional predictive ordinates (CPO) \citep{lewis2014posterior}, and the pointwise predictive density (\LPPD) \citep{vehtari2017practical}.
Finally, the nested sampling (NS) method sits somewhere in between the single- and multiple-chain categories as it requires simulations from multiple short MCMC runs \citep{skilling2004nested,skilling2006nested,maturana2018nested}.

The final set of methods all require multiple chains, which are ``heated'' with a heating parameter that interpolates between the posterior distribution and some other distribution.
For the path sampling \citep{ogata1989monte,gelman1998simulating,lartillot2006computing,friel2008marginal,baele2012improving} and stepping stone (SS) methods \citep{xie2010improving}, the power posterior path links the posterior to the prior distribution.
\citet{fan2010choosing} proposed the generalized stepping stone (GSS) method in which the path is defined between the posterior and a reference distribution, hence avoiding issues associated with sampling from vague priors.

A number of the above methods have been previously applied to phylogenetics, including all power posterior approaches, the harmonic mean, and conditional predictive ordinates.
In phylogenetics, path sampling and stepping stone are currently the most widely used methods, and are included in popular inference programs like BEAST \citep{drummond2012bayesian} and MrBayes \citep{ronquist2012mrbayes}.

\subsubsection*{Laplus}
The Laplace approximation \citep{tierney1986accurate,kass1995bayes} replaces the true log-posterior distribution with a multivariate normal distribution.
%EM Very minor point. The hessian is in general the matrix of second derivatives. Wouldn't it be more informative here to just call this the inverse of the observed information matrix?
%AM I don't think that information matrices are any simpler for our intended audience than Hessians, and possibly are more foreign
The mean is taken to be the joint posterior mode ($\boldsymbol{\tilde{\theta}} = (\tilde{\theta}_1,\tilde{\theta}_2, \dots ,\tilde{\theta}_{2S-3})$, and the covariance matrix is taken to be  the inverse of the observed information matrix of $l(\boldsymbol{\theta}) = \log(p(D|\boldsymbol{\theta}, \tau)\,p(\boldsymbol{\theta} \mid \tau))$ evaluated at $\boldsymbol{\tilde{\theta}}$.
Previous studies have approximated the likelihood surface of phylogenies using multivariate normal distributions \citep{thorne1998estimating,guindon2010bayesian}, including the use of parameter transformations to account for positivity and skew \citep{reis2011approximate}.
However, the posterior distribution of branch lengths may have its mode at 0 in some dimensions, which is not a shape that can be attained by any transformation of a normal distribution.
In related work, the conditional posterior distribution of single branch lengths has been approximated with a gamma distribution, which can accommodate the zero mode, enabling independence sampling \citep{aberer2015efficient}.

We depart from the aforementioned approaches and introduce a novel framework to approximate the joint posterior distribution on branch lengths.
For simplicity, in all cases we assume that \textit{a posteriori} branch lengths are independent.
This is obviously not true in practice, but we find that posterior correlations are often quite small, and that our independence assumption works well.
This assumption also greatly reduces the computational burden by allowing us to sidestep computing every second partial derivative.

Our ``Laplus'' approximation then takes the maximum \textit{a posteriori} (MAP) vector of branch lengths $\boldsymbol{\tilde{\theta}}$ and the vector of second derivatives $\left(\frac{\partial^2l}{\partial \theta_1^2}, \frac{\partial^2l}{\partial \theta_2^2}, \dots, \frac{\partial^2l}{\partial \theta_{2S-3}^2}\right)$ and finds the parameters of our approximating distributions for each branch, $\boldsymbol{\phi}_i$, by matching modes and second derivatives of the approximating and posterior distributions of branch lengths.
Unlike the method of moments and maximum likelihood estimation, our approach is fast as it does not require a set of samples to estimate the parameters of the distribution.
We consider three distributions for approximating the marginal posteriors of branch lengths: lognormal, gamma, and beta$'$ (\IE\ beta prime).
The general procedure for the Laplus approximations is similar regardless of what distribution (\IE\ the choice of $q$ in $q(x; \boldsymbol{\phi}_{i})$) is chosen to approximate the posterior, and is written here algorithmically:

\begin{enumerate}
  \item Find the (joint) MAP branch lengths, $\boldsymbol{\tilde{\theta}} = (\tilde{\theta}_1,\tilde{\theta}_2, \dots ,\tilde{\theta}_{2S-3})$
  \item For $i = 1,\dots, 2S-3$
    \begin{enumerate}
      \item Compute $\frac{\partial^2l}{\partial \theta_i^2}$, the second derivative of the log unnormalized posterior with respect to the $i^{\textnormal{th}}$ branch
      \item Find parameters of $\boldsymbol{\phi}_{i}$ by solving
\begin{align*}
\frac{d^2}{dx^2}\log(q(x; \boldsymbol{\phi}_{i})) &=\frac{\partial^2l}{\partial \theta_i^2}\Big|_{\theta_i = \tilde{\theta}_i}, \\
\text{mode}(q(x; \boldsymbol{\phi}_{i})) &= \tilde{\theta}_i
	\end{align*}
      \item Catch exceptions
    \end{enumerate}
  \item Compute the marginal likelihood as $\hat{p}_{\text{Laplus}}(D \mid \tau) = \frac{p(D \mid \boldsymbol{\tilde{\theta}},\tau)p(\boldsymbol{\tilde{\theta}} \mid \tau)}{\prod_i q(\tilde{\theta}_i ; \boldsymbol{\phi}_{i})}$.
\end{enumerate}

Exceptions occur when elements of $\boldsymbol{\phi}_{i}$ are outside of the domain of support, when the second derivative is nonnegative (so the posterior has a mode at 0), or when elements of $\boldsymbol{\phi}_{i}$ are otherwise suspect (such as producing particularly high-variance distributions with very short branches).
Exceptions and their handling depend on the distributional kernel (choice of $q$), and we defer a full discussion of this to the supplementary material.

\subsubsection*{Variational inference}
The main idea behind variational inference is to transform posterior approximation into an optimization problem using a family of approximate densities.
The aim is to find the member of that family with the minimum Kullback-Leibler (KL) divergence to the posterior distribution of interest:
\[
\boldsymbol{\phi}^{*} = \argmin_{\boldsymbol{\phi} \in \boldsymbol{\Phi}}  \mathrm{KL}(q(\boldsymbol{\theta}; \boldsymbol{\phi}) \parallel p(\boldsymbol{\theta} \mid D, \tau)),
\]
where $q(\boldsymbol{\theta}; \boldsymbol{\phi})$ is the variational distribution parametrized by a vector $\boldsymbol{\phi} \in \boldsymbol{\Phi}$ and  KL is defined as
\[
\mathrm{KL}(q \parallel p)=\int_{\boldsymbol{\theta}} q(\boldsymbol{\theta}; \boldsymbol{\phi}) \log \frac{q(\boldsymbol{\theta}; \boldsymbol{\phi})}{p(\boldsymbol{\theta} \mid D, \tau)}.
\]
To minimize the KL divergence, we first rewrite the KL equation:
$$
\begin{aligned}
\mathrm{KL}(q(\boldsymbol{\theta}; \boldsymbol{\phi}) \parallel p(\boldsymbol{\theta} \mid D, \tau)) &= \mathop{\mathbb{E}}[\log q(\boldsymbol{\theta}; \boldsymbol{\phi})] - \mathop{\mathbb{E}}[\log p(\boldsymbol{\theta} \mid D, \tau)] \\
  & = \mathop{\mathbb{E}}[\log q(\boldsymbol{\theta}; \boldsymbol{\phi})] - \mathop{\mathbb{E}}[\log p(\boldsymbol{\theta}, D \mid \tau)] + \log p(D \mid \tau),
\end{aligned}
$$
where the expectations are taken with respect to the variational distribution $q$.
The third term $\log p(D \mid \tau)$ on the right hand side of the last equality is a constant with respect to the variational distribution so it can be ignored for the purpose of the minimization.
After switching the sign of the other two terms, the minimization problem can be framed as a maximization problem of the function
\[
\textrm{ELBO}(\boldsymbol{\phi}) = \mathop{\mathbb{E}}[\log p(\boldsymbol{\theta}, D \mid \tau)] - \mathop{\mathbb{E}}[\log q(\boldsymbol{\theta}; \boldsymbol{\phi})].
\]
The ELBO is easier to calculate than the KL divergence as it does not involve computing the intractable posterior normalisation term $p(D \mid \tau)$.
The ELBO gives a lower bound of the marginal likelihood, the very measure we are interested in estimating here.
Here we use the ELBO estimate  $\hat{p}_{\textrm{ELBO}}(D \mid \tau) := \max_{\boldsymbol{\phi} \in \boldsymbol{\Phi}}\textrm{ELBO}(\boldsymbol{\phi})$ to approximate the marginal likelihood of a topology.

We used a Gaussian variational mean-field approximation applied to log-transformed branch lengths to ensure that the variational distribution stays within the support of the posterior.
The mean-field approximation assumes complete factorisation of the distribution over each of the $2S-3$ branch length variables and each factor is governed by its own variational parameters $\boldsymbol{\phi}_i$:
\[
q(\theta_1, \dots, \theta_{2S-3}; \boldsymbol{\phi}) = \prod_{i=1}^{2S-3} q(\theta_i; \boldsymbol{\phi}_i),
\]
where $q(\theta_i; \boldsymbol{\phi}_i)$ is  a log-normal density and $\boldsymbol{\phi}_i = (\mu_i, \sigma_i)$.
As in the Laplus approximation, this model also assumes that there is no correlation between branches.

The variational parameters are estimated using stochastic gradient ascent using a black box approach \citep{ranganath2014black} similar to the algorithm implemented in Stan \citep{kucukelbir2015automatic}.

\subsubsection*{Importance sampling}
The Laplus and variational Bayes approximations of the marginal likelihood are fast, but in practice the approximate posterior does not always match the posterior of interest well.
Since these methods rely on independent univariate probability distributions (\EG\ gamma, normal, etc), samples can be efficiently drawn from the approximate posterior distributions.
We thus also used importance sampling to reduce the bias of the Laplus and variational Bayes methods using the approximate posterior distribution as the importance instrument distribution.

The importance sampling estimate of $p(D \mid \tau)$ using an approximate normalized probability distribution (instrument distribution) $g$ is
\[
\hat{p}_{\textrm{IS}}(D \mid \tau) = \frac{1}{N} \sum_{i=1}^N \frac{p(D \mid \tilde{\boldsymbol{\theta}}_i, \tau) p(\tilde{\boldsymbol{\theta}}_i \mid \tau)}{g(\tilde{\boldsymbol{\theta}}_i)}\text{, where } \tilde{\boldsymbol{\theta}}_i \sim g(\boldsymbol{\theta}).
\]

\subsection*{Benchmarks}
We benchmark the 19 methods for estimating fixed-tree marginal phylogenetic likelihood on 5 empirical datasets from a suite of standard test datasets \citep{lakner2008efficiency,hohna2011guided,larget2013estimation,whidden2015quantifying}, which we call DS1 through DS5.
These datasets vary from 25 to 50 taxa, with alignment number of sites ranging from 378 to 2520.
Instead of focusing primarily on the accuracy of the estimate of the single-tree marginal likelihoods, we focus on the approximate posterior of topologies we obtain by applying our marginal likelihood methods to each and normalizing the result as described below.
We take measures of the goodness of these posteriors that directly address approximation error in quantities of interest, namely the posterior probabilities of topologies and the probabilities of tree splits.
These are compelling choices because Bayesian phylogenetic inference is not performed to answer the question ``what is the marginal likelihood of this topology" but rather to quantify support for evolutionary relationships/hypotheses.
We note that the posterior of trees is also useful in other contexts, such as examining the information content of a dataset \citep{lewis2016estimating}.

To compare marginal likelihood methods' accuracy and precision, we need to establish a ground truth for $p(\tau_i \mid D)$ for each tree topology $\tau_i$.
To do this, we use the extensive runs (called golden runs) of MrBayes from \citet{whidden2015quantifying}, which consist of 10 chains run for 1 billion generations each (subsampled every 1000 generations), with 25\% discarded as burnin and all chains pooled when computing posterior summaries.
This results in 7.5 million MCMC samples from 7.5 billion generations, with common diagnostics showing convergence of the chains.
The credible sets contain between 5 and 1,141,881 topologies.
For datasets DS1 to DS4, we run each of the 19 methods for calculating marginal likelihoods on every tree in the 95\% posterior credible set.
DS5 has a credible set that is too large (over one million topologies), so we consider only the 1000 most probable trees from this dataset.
The only input for each of the 19 methods from the golden runs is the tree topology without branch lengths.
In the Golden runs,  \texttt{MrBayes} was set up to use a uniform prior for topologies and independent exponential priors with mean 0.1 for the branch lengths.

After arriving at a set of trees for each benchmark dataset, we renormalize \texttt{MrBayes} posterior probabilities so that they sum to one over the selected trees: $\sum_i P(\tau_i \mid D) = 1$.
We assume these probabilities form the true posterior mass function of tree topologies and measure accuracy with respect to this function.
We use the Bayes rule to convert our approximations of the marginal likelihood to the posterior probability:
\[
\hat{p}(\tau_i \mid D) = \frac{\hat{p}(D \mid \tau_i) p(\tau_i)}{\sum_j \hat{p}(D \mid \tau_j) p(\tau_j)} =
 \frac{\hat{p}(D \mid \tau_i)}{\sum_j \hat{p}(D \mid \tau_j)},
\]
where the last equality holds because we assumed the uniform prior over the tree topologies.
The marginal likelihood estimations were replicated 10 times for each combination of method and dataset, allowing us to derive the standard deviation of the marginal likelihood estimates.

We employ two different measures to determine closeness of an approximate posterior to the golden run posterior.
Since many questions in phylogenetics concern the probabilities of individual splits, we consider the error in their estimated posterior probabilities.
%MF It might be obvious from the context but we use S for the number of sequences and the number of splits. It could be replaced with B
We calculate the root mean-squared deviation (RMSD) of the probabilities of splits, computed as $\text{RMSD} = \sqrt{\frac{1}{S} \sum_i (\hat{f}(s_i) - f(s_i))^2}$, where $s_i$ is a split (or bipartition) and $S$ the number of splits in the tree topology set.
The probabilities of a split are given by $f(s_i) = \sum_j p(\tau_j \mid D)\,\mathbbm{1}_{s_i \in \tau_j}$ and $\hat{f}(s_i) = \sum_j \hat{p}(\tau_j \mid D)\,\mathbbm{1}_{s_i \in \tau_j}$, that is, they are the sums of posterior probabilities of the topologies that contain that split.
To assess how well the posterior probabilities of topologies are estimated, we use the Kullback-Leibler (KL) divergence from $\hat{\mathbf{p}}= (\hat{p}(\tau_1 \mid D), \dots, \hat{p}(\tau_N \mid D))$ to $\mathbf{p}= (p(\tau_1 \mid D), \dots, p(\tau_N \mid D))$, where $N$ is the number of unique topologies in the 95\% posterior credible set of the golden run.
This is computed as as $\mathrm{KL}(\mathbf{p} \parallel \hat{\mathbf{p}})=\sum_i p(\tau_i \mid D) \log \frac{p(\tau_i \mid D)}{\hat{p}(\tau_i \mid D)}$.

Given that these 19 marginal likelihood calculation methods vary widely in their computational efficiency, we also seek to benchmark the speed of the methods.
As our measure of speed, we take the average time (per dataset) required to compute the marginal likelihood of a topology.
The speed of these methods depends on a number of dataset-specific features (including on the size of the dataset and the number of phylogenies in the credible set), on run-time decisions (such as the number of MCMC iterations), and on the code that implements them.
By incorporating multiple datasets (to average over dataset-specific effects) and implementing the methods in a single package (to control for run-time and implementation-specific effects), we are able to examine the general tradeoff between speed and accuracy, and highlight the use-cases we think the methods are suited for.

Every method was implemented within the phylogenetic package \texttt{physher} \citep{fourment2014novel} (\url{https://github.com/4ment/physher}) and we used the same priors as in the golden runs of MrBayes.
Datasets and scripts used in this study are available from \url{https://github.com/4ment/marginal-experiments/}.
We note that this study used a single-threaded version of physher, leaving much room to improve the speed of these embarrassingly parallelizable algorithms.
All analyses were run on Intel Xeon E5-26972.60GHz processors running CentOS release 6.1 with 244 GB of RAM.

\section*{Results}
\subsection*{Accuracy and precision}
\subsubsection*{RMSD}
\begin{figure}[H]
\begin{center}
\includegraphics[width=0.8\linewidth]{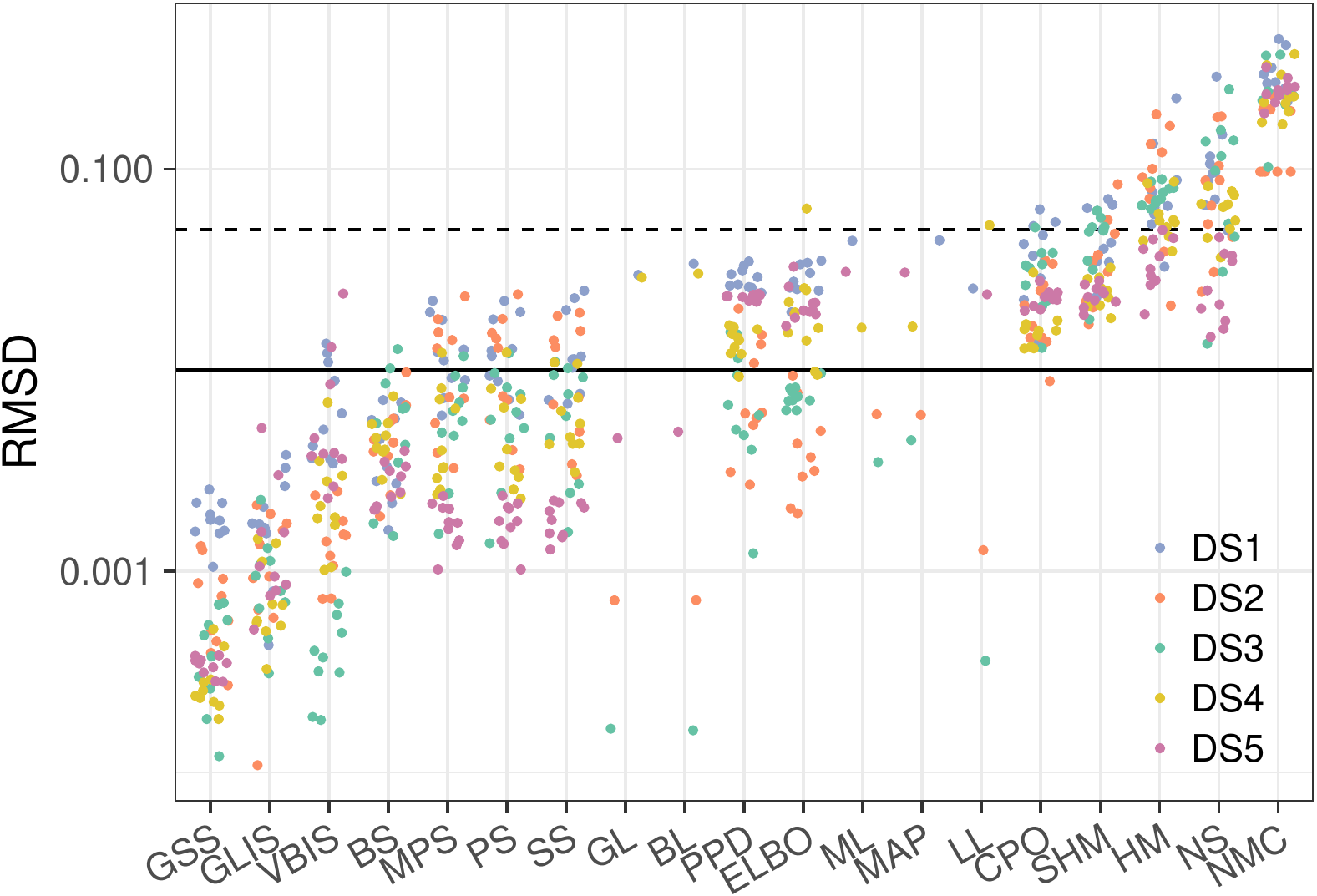}
\end{center}
\caption{Average split posterior RMSD for 10 replicate runs of each dataset. \LL, \GL,  BL, MAP, and ML are deterministic and therefore only one replicate is shown. The horizontal dashed and solid lines depict RMSDs of 0.05 and 0.01, respectively.}
\label{fig:rmsd_by_ds}
\end{figure}

\begin{figure}[H]
\begin{center}
\includegraphics[width=0.8\linewidth]{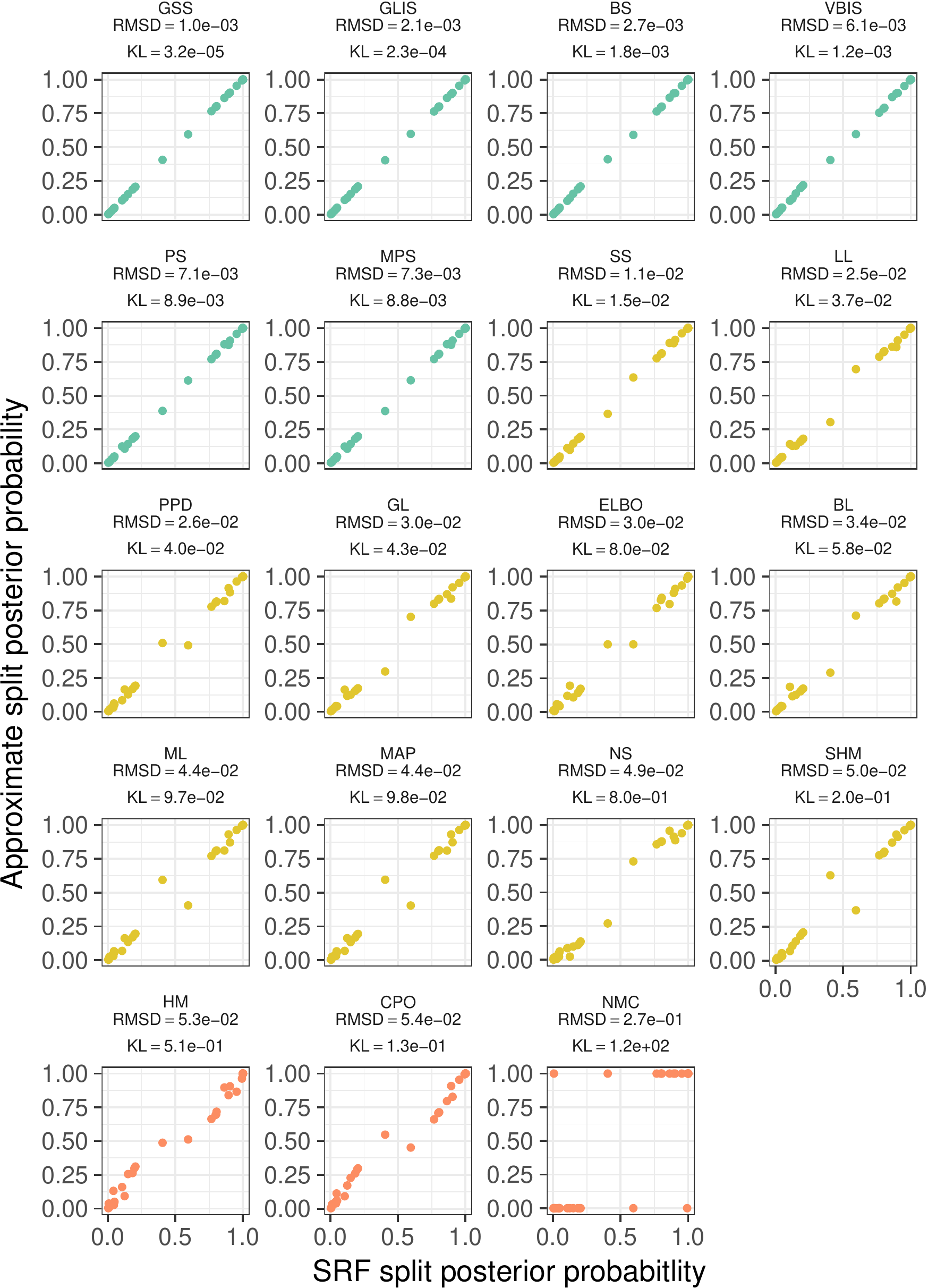}
\end{center}
\caption{The posterior probabilities of all the splits observed in DS1 for a single replicate. MrBayes posteriors are plotted on the x-axis versus the denoted approximation on the y-axis. The line $y=x$ is provided for ease of interpretation, and points are colored by the thresholds we discuss: RMSD $<$ 0.01 is a good approximation (green), 0.01 $\leq$ RMSD $<$ 0.05 is a potentially acceptable approximation (yellow), and RMSD $\geq$ 0.05 is poor (red). Panels are ordered by RMSD in increasing order.}
\label{fig:split_probs_1}
\end{figure}

\begin{figure}[H]
\begin{center}
\includegraphics[width=0.8\linewidth]{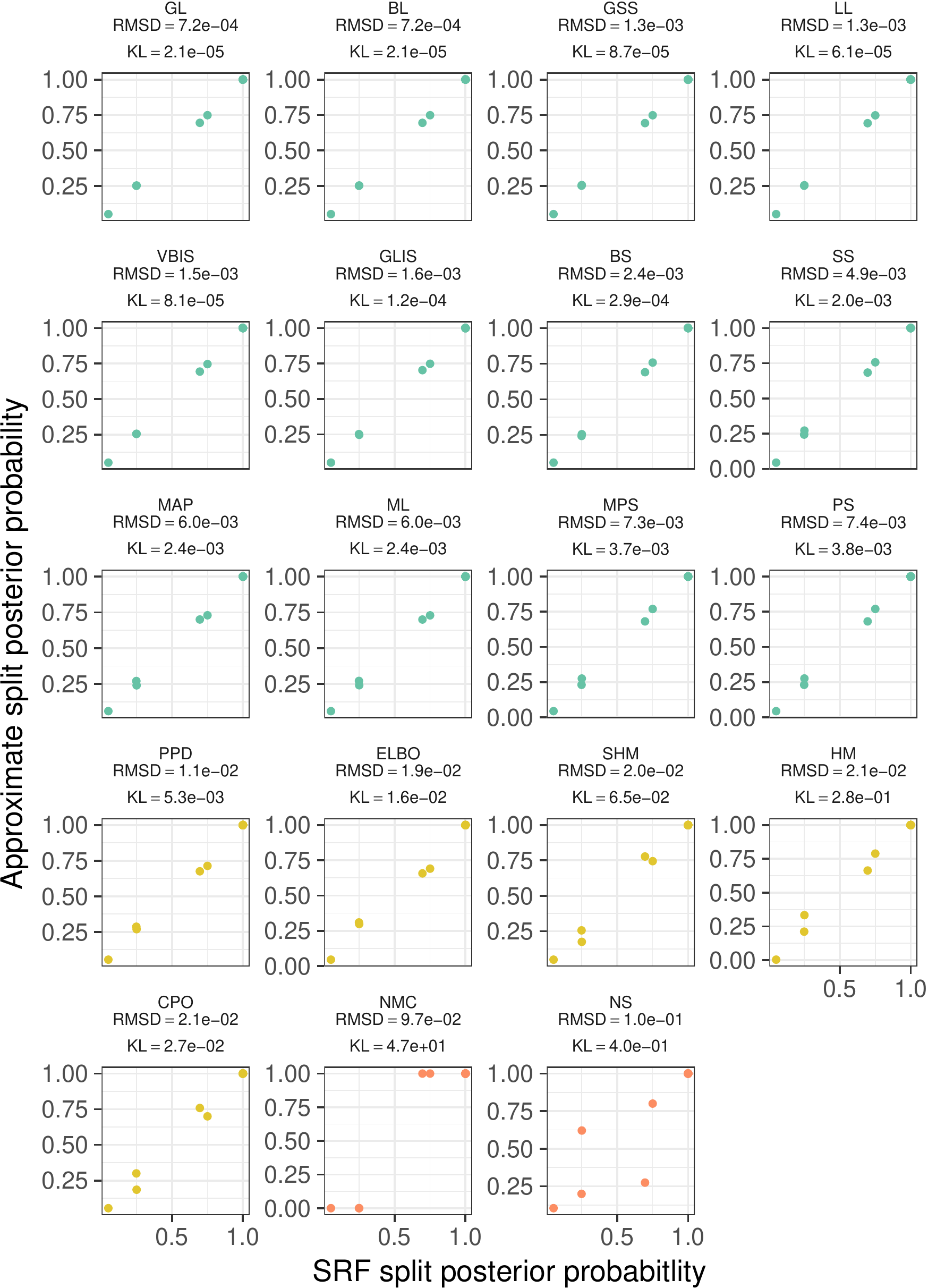}
\end{center}
\caption{The posterior probabilities of all the splits observed in DS2 for a single replicate. MrBayes posteriors are plotted on the x-axis versus the denoted approximation on the y-axis. The line $y=x$ is provided for ease of interpretation, and points are colored by the thresholds we discuss: RMSD $<$ 0.01 is a good approximation (green), 0.01 $\leq$ RMSD $<$ 0.05 is a potentially acceptable approximation (yellow), and RMSD $\geq$ 0.05 is poor (red). Panels are ordered by RMSD in increasing order.}
\label{fig:split_probs_2}
\end{figure}

\begin{figure}[H]
\begin{center}
\includegraphics[width=0.8\linewidth]{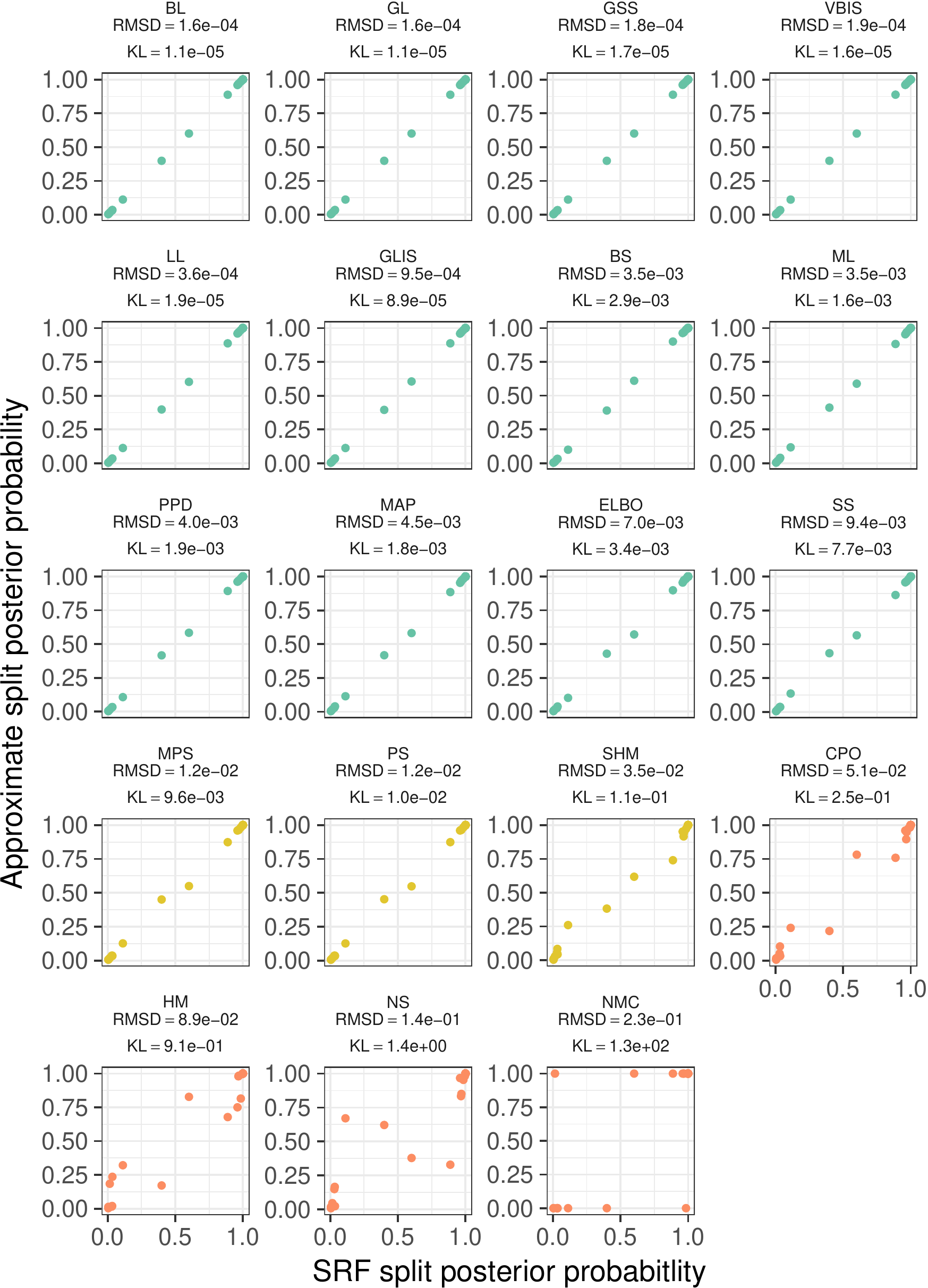}
\end{center}
\caption{The posterior probabilities of all the splits observed in DS3 for a single replicate. MrBayes posteriors are plotted on the x-axis versus the denoted approximation on the y-axis. The line $y=x$ is provided for ease of interpretation, and points are colored by the thresholds we discuss: RMSD $<$ 0.01 is a good approximation (green), 0.01 $\leq$ RMSD $<$ 0.05 is a potentially acceptable approximation (yellow), and RMSD $\geq$ 0.05 is poor (red). Panels are ordered by RMSD in increasing order.}
\label{fig:split_probs_3}
\end{figure}

\begin{figure}[H]
\begin{center}
\includegraphics[width=0.8\linewidth]{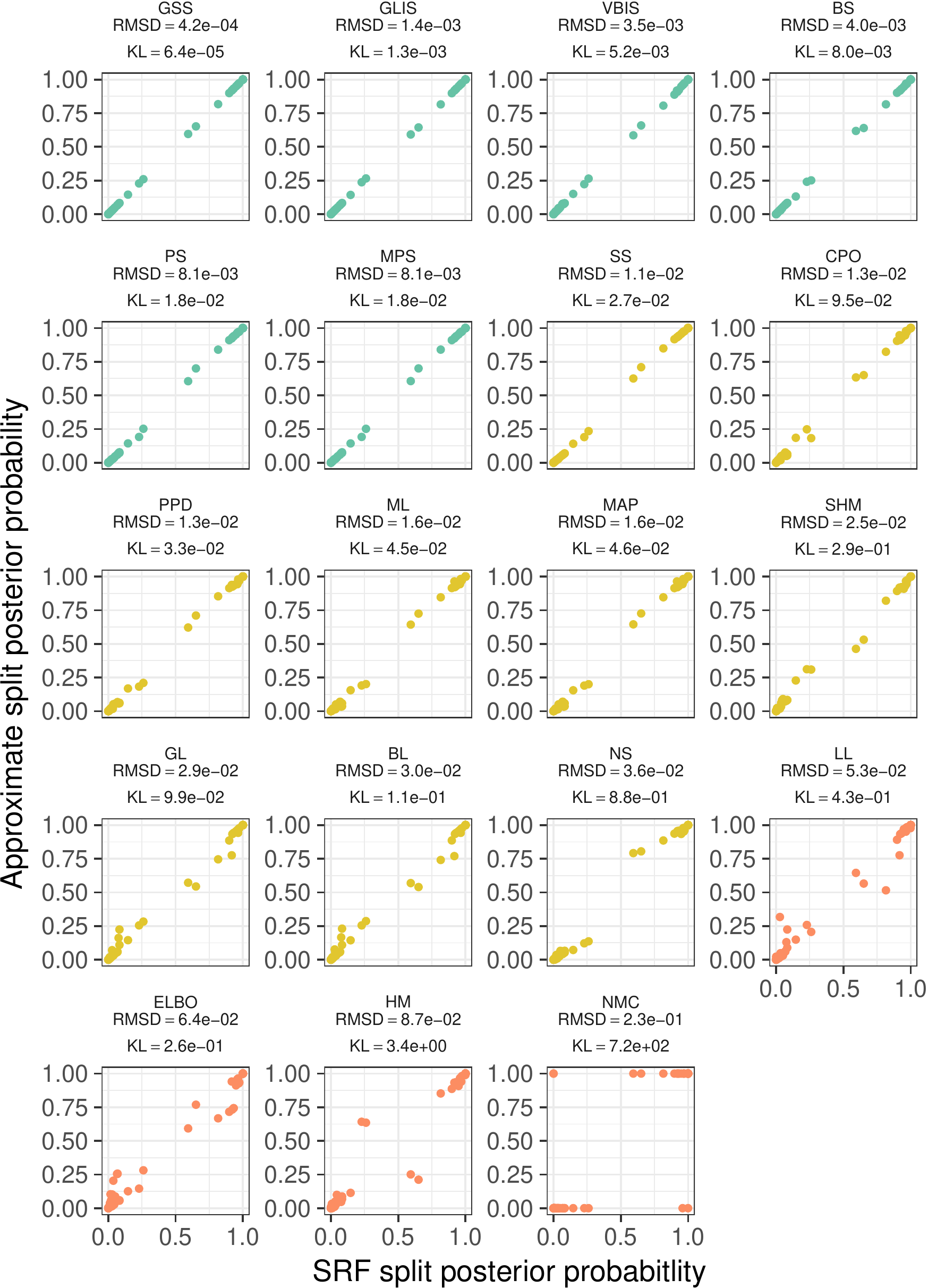}
\end{center}
\caption{The posterior probabilities of all the splits observed in DS4 for a single replicate. MrBayes posteriors are plotted on the x-axis versus the denoted approximation on the y-axis. The line $y=x$ is provided for ease of interpretation, and points are colored by the thresholds we discuss: RMSD $<$ 0.01 is a good approximation (green), 0.01 $\leq$ RMSD $<$ 0.05 is a potentially acceptable approximation (yellow), and RMSD $\geq$ 0.05 is poor (red). Panels are ordered by RMSD in increasing order.}
\label{fig:split_probs_4}
\end{figure}

%EM It's rather surprising to expect results and then get what looks like a motivation for methods. Move this bit up into methods.
%MF Maybe we should talk about that before making big changes. I remember that we talked about statements that could  either fit in the methods or results.
When comparing multiple replicate MCMC analyses (multiple runs), a standard metric in phylogenetics is the average standard deviation of split frequencies (ASDSF).
Typically an ASDSF below 0.01 is taken to be evidence that two MCMC analyses are sampling the same distribution.
We use the related (but stricter) RMSD as our measure of approximation error (Figure \ref{fig:rmsd_by_ds}).
By considering the plots of split probabilities organized by their RMSD, (Figure~\ref{fig:split_probs}, Supplementary Figures \ref{fig:split_probs_1}, \ref{fig:split_probs_2}, \ref{fig:split_probs_3}, and \ref{fig:split_probs_4}), we developed two cutoffs for RMSD to classify method performance.
We call methods with RMSD less than 0.01 to be in ``good'' agreement with ground truth, while we say that methods with RMSD between 0.01 and 0.05 are in ``acceptable'' agreement.
RMSD above 0.05 indicates substantial disagreement between ground truth and estimates.
%EM Here's what I consider our first real results statement, which is that the accuracy of the results is consistent across 5 data sets.
Most of the 19 methods' estimates fall within these categories consistently across the 5 datasets.
MAP, ML, \GL\ and BL span the boundary between good and acceptable, while \LL\ spans all three categories. Recall that all methods abbreviations are in Table~\ref{tab:method_names}.

\subsubsection*{KL divergence}
\begin{figure}[H]
\begin{center}
\includegraphics[width=0.8\linewidth]{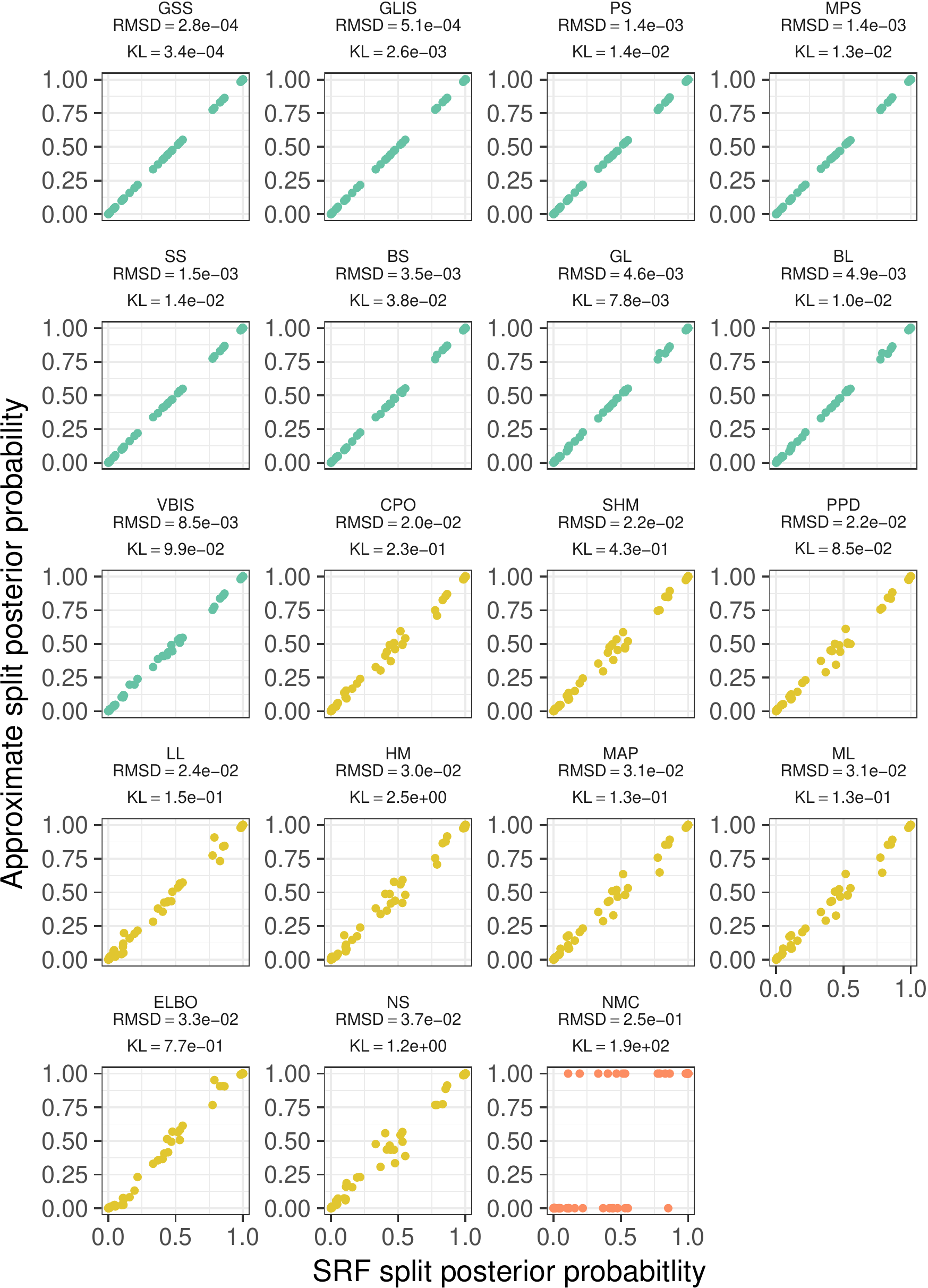}
\end{center}
\caption{The posterior probabilities of all the splits observed in DS5 for a single replicate. MrBayes posteriors are plotted on the x-axis versus the denoted approximation on the y-axis. Points are colored by the thresholds we discuss: RMSD $<$ 0.01 is a good approximation (green), 0.01 $\leq$ RMSD $<$ 0.05 is a potentially acceptable approximation (yellow), and RMSD $\geq$ 0.05 is poor (red). Panels are ordered by RMSD in increasing order.}
\label{fig:split_probs}
\end{figure}

\begin{figure}[H]
\begin{center}
\includegraphics[width=0.8\linewidth]{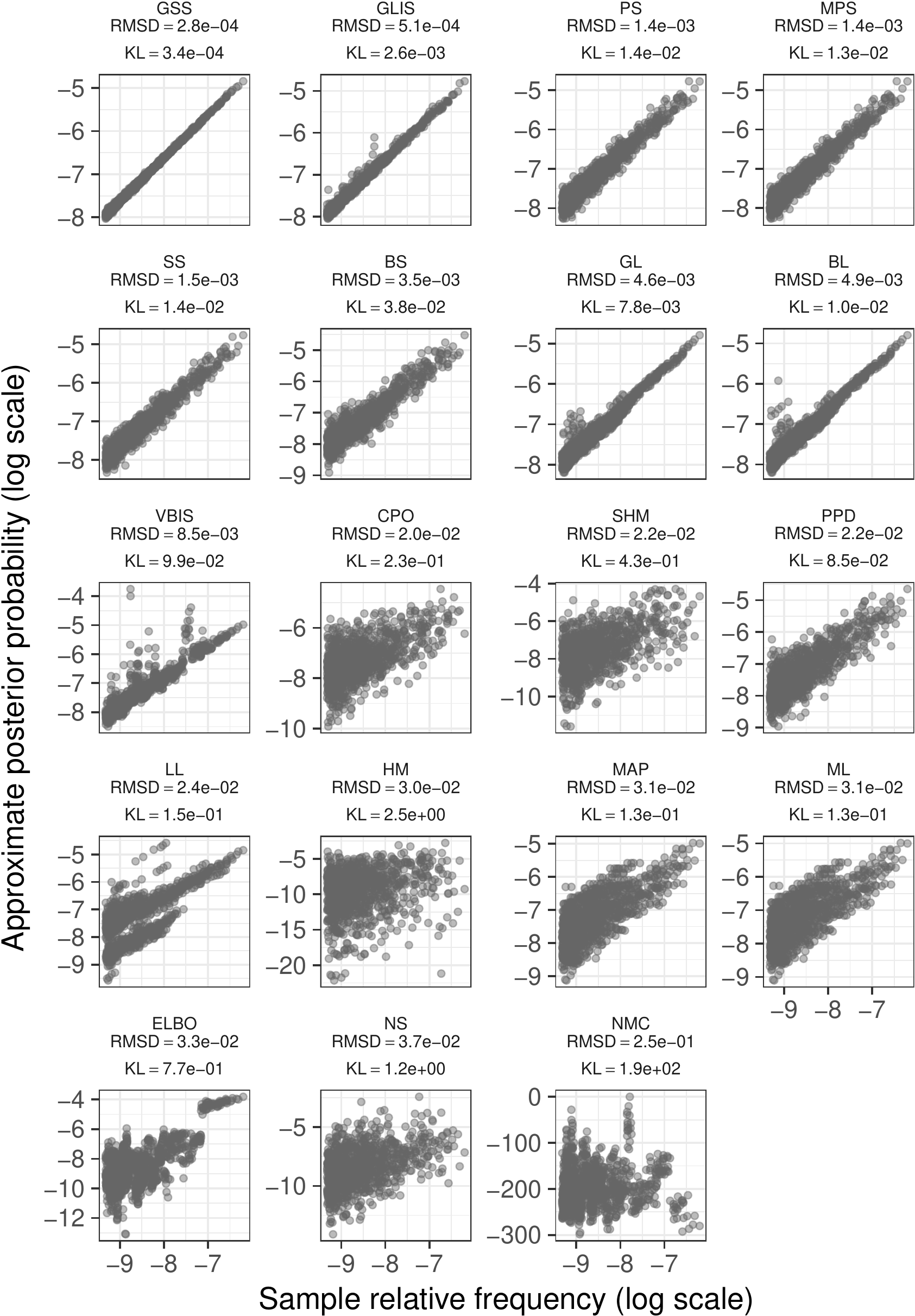}
\end{center}
\caption{The approximate posterior probabilities of the topologies in DS5 versus the ground truth posterior probabilities from MrBayes, plotted on the log scale for clarity. The rank-ordering of the methods is closest to average for DS5. Results are for a single run of each method. Panels are ordered by RMSD in increasing order.}
\label{fig:approx_vs_srf}
\end{figure}

Broadly speaking, there is concordance between the performance of approximations whether measured by KL divergence or RMSD (Figures \ref{fig:split_probs}, \ref{fig:approx_vs_srf}).
This is reassuring, as a good approximation should estimate the marginal likelihoods well, which should result in good approximations to the posterior, and thus good estimation of the split probabilities.
We also find that the methods do a better job approximating the marginal likelihood of more probable trees than less probable trees (seen as triangular shapes of scatter points in Figure~\ref{fig:approx_vs_srf}).
However, even methods that lead to notable scatter between truth and approximation, such as \LPPD, can yield quite good estimates of the probabilities of splits.
Additionally, if the only quantity of interest is the 50\% majority-rule consensus tree, then even methods that estimate the marginal likelihood quite poorly can lead to reasonable trees (Figure~\ref{fig:consensus_trees}).
To get the same consensus tree, a method must merely place the same splits in the upper 50\% range of posterior probability, so this measure can hide a substantial amount of variability in the estimated marginal likelihoods.

\subsection*{Speed}
\begin{figure}[H]
\begin{center}
\includegraphics[width=0.8\linewidth]{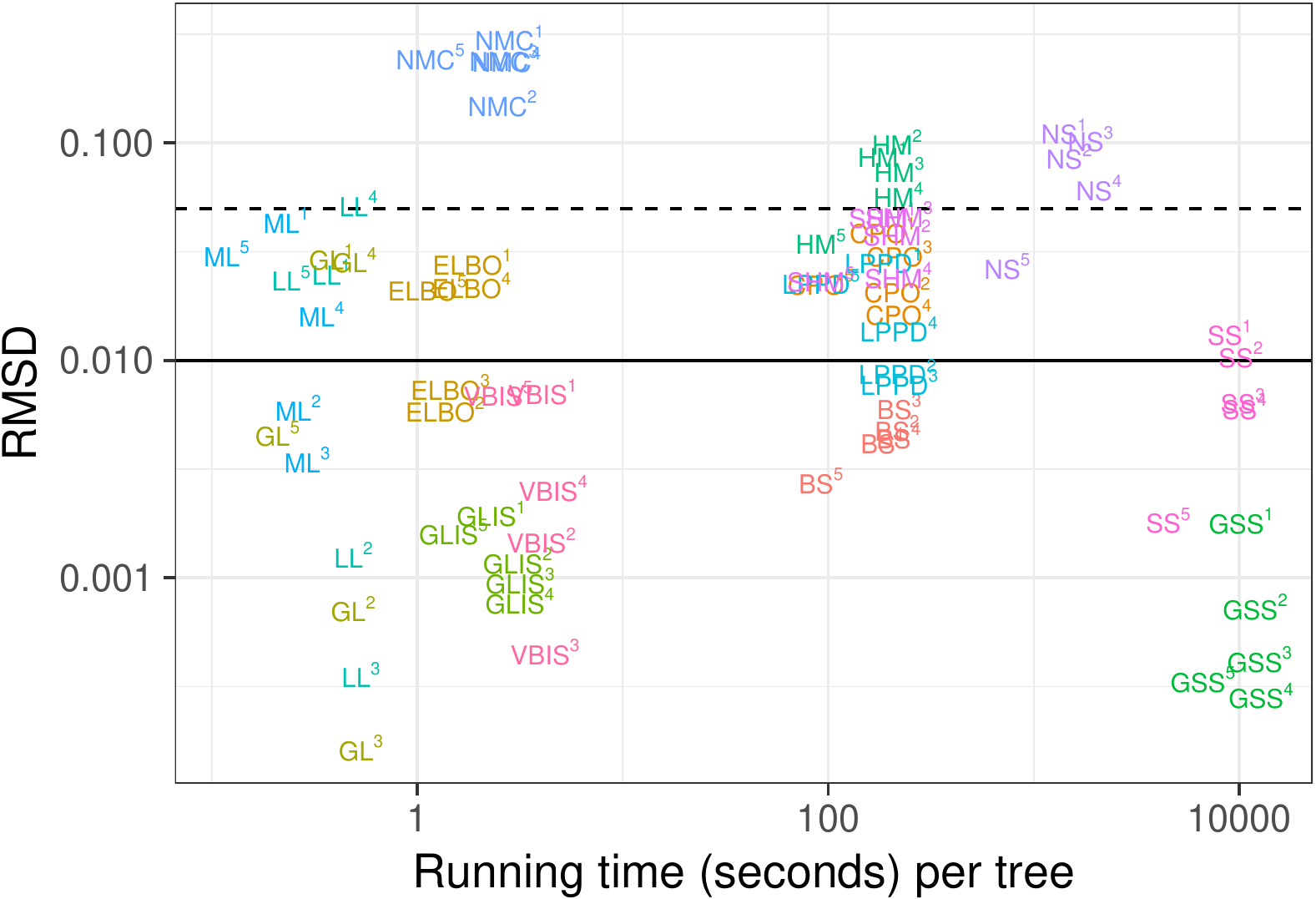}
\end{center}
\caption{
Average RMSD of splits in the approximate posterior against running time. Text denotes method used, while superscripts label applications to individual datasets. Four methods are omitted for visual clarity: MAP is essentially identical to ML, BL is nearly identical to \GL, and \PS\ and \PStwo\ are both similar to \SStone. The horizontal dashed and solid lines depict RMSDs of 0.01 and 0.05 respectively. The RMSD is calculated using the average marginal likelihood of each tree from each of 10 replicate analyses. The running time is calculated using the average running time of each tree from each of 10 replicate analyses.}
\label{fig:time_vs_accuracy}
\end{figure}

Fast methods can give accurate results, while slow methods need not be accurate (Figure~\ref{fig:time_vs_accuracy}).
Indeed, \GL\ is very fast to compute and gives good results, \GLIS\ is only slightly slower and gives excellent results, while NS is slow to compute and gives rather bad results for this problem.

Method speed is primarily determined by the amount of sampling performed by the method: the more sampling required by a method, the slower it is.
The fastest methods are deterministic and do not perform sampling at all, with MAP and ML being the fastest of the 19, requiring only optimization.
There is a minor added computational cost of calculating additional derivatives of the phylogenetic likelihood function (here purely the derivatives with respect to branch lengths) in the case of the Laplus approximations.
The calculation of the ELBO is slightly slower due to the cost of optimizing the variational parameters through stochastic gradient ascent.
The next jump in speed is to methods that perform importance sampling.
The single-chain methods are very consistent in time requirements since the computation time is largely dominated by the MCMC. They are notably slower than the importance sampling methods, because MCMC here used one million samples per tree, while we use 10,000 for importance sampling.
The slowest methods require running multiple MCMC chains, and aside from \GSS\ time requirements are essentially identical between these methods.
We used 50 power posteriors in our analysis of stepping stone and path sampling methods, and as expected we find that they are very nearly 50 times slower than the single-chain methods.
The consistency of the number of chains and the time requirement of the method clearly demonstrates that the largest computational effort is in the MCMC.
It is worth noting, though, that after an MCMC analysis has run (power posteriors or single chains), any appropriate method can be used to post-process the chains and calculate the marginal likelihood, as MrBayes does with arithmetic and harmonic means.
As an implementation detail of this study, every single-chain method uses the same MCMC samples to estimate the marginal likelihood and similarly, the power posterior-based methods use the same power posterior samples.

\subsection*{Monte Carlo error}
\begin{figure}[H]
\begin{center}
\includegraphics[width=0.8\linewidth]{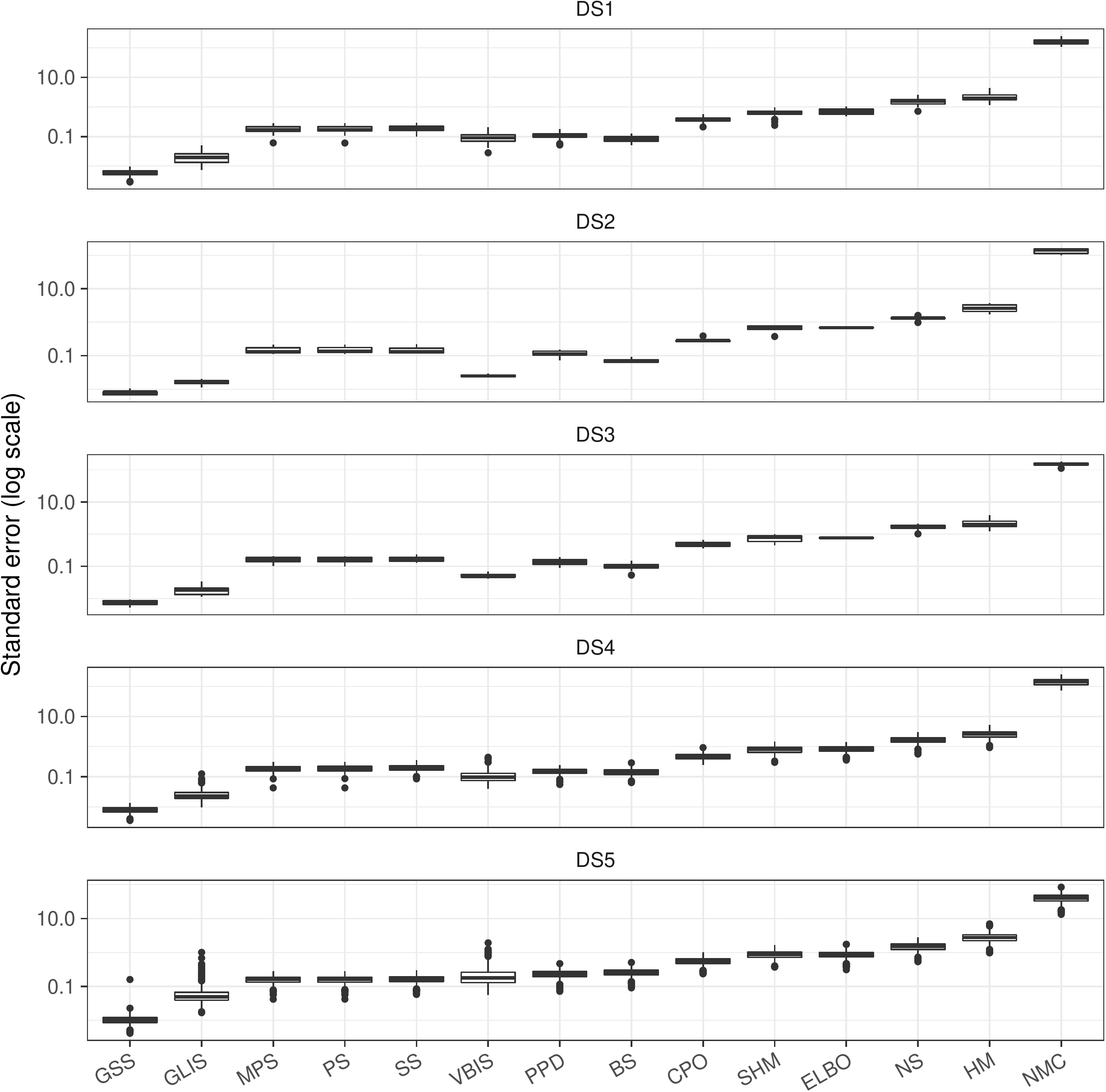}
\end{center}
\caption{Standard error of the Monte-Carlo-based estimators. Each point represents the standard error of an individual tree across the 10 replicate analyses for each estimator.}
\label{fig:estimator_variability}
\end{figure}

No method to estimate the posterior probability of a tree is without sources of error.
Monte Carlo error is a feature of all of sampling-based methods we benchmarked, including the methods using at least one MCMC chain and importance sampling methods (marked by asterisks in Table \ref{tab:method_names}).
For these methods, and the variational approach (which uses stochastic optimization with noisy gradient estimates and thus also has inter-run variability) we ran 10 replicate analyses (Figure~\ref{fig:estimator_variability}).
Interestingly, we find that the inter-run variability of the methods is correlated with the goodness of the estimates (and hence the rank-orderings of the methods are similar in Figure~\ref{fig:estimator_variability} and Figure~\ref{fig:rmsd_by_ds}).
%EM I suggest moving this up into the methods, if it's not already. I see this in the figure legends of course.
In discussing how well the methods approximate the posterior distribution of trees, to diminish the effects of Monte Carlo error, we use the average estimated marginal likelihood across the replicate analyses.

\subsection*{Summary trees}
\begin{figure}[H]
\begin{center}
\includegraphics[width=0.8\linewidth]{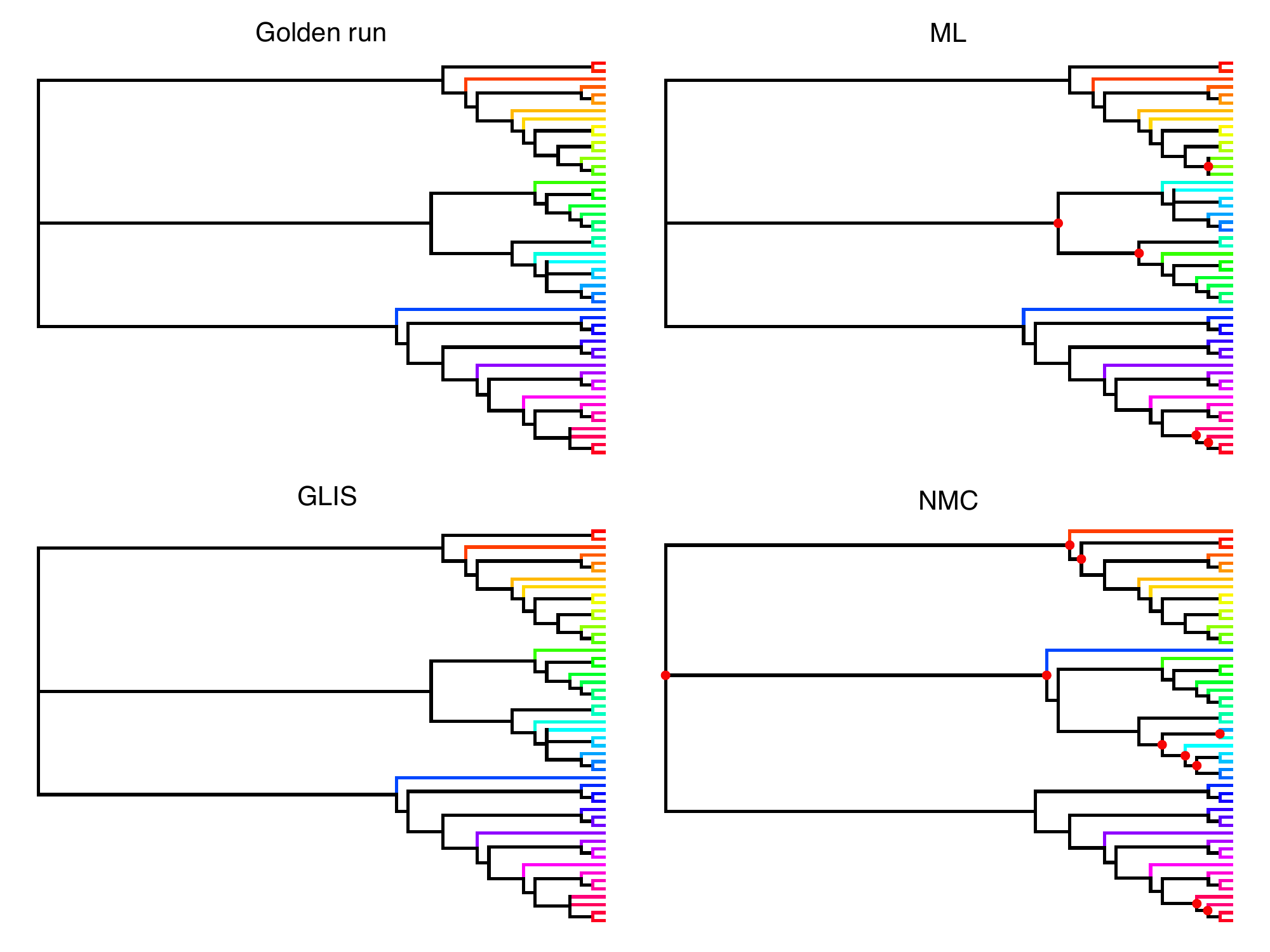}
\end{center}
\caption{Majority rule consensus trees DS5 based on four sources for posterior probabilities of trees. Each taxon is assigned a unique color and the branch leading to that taxon is colored the same in all 4 trees to show differences. The golden run and \GLIS\ trees are identical, while the tree for ML has a Robinson-Foulds distance of 4 to those trees and the tree for NMC a distance 14 (and 10 from the ML tree). Nodes with red circles denote parts of the tree different from the golden run tree.}
\label{fig:consensus_trees}
\end{figure}

The accuracy of summary trees was correlated as expected with the accuracy of the posterior estimate on splits (Figure~\ref{fig:consensus_trees}).
% When presenting the results of a phylogenetic study, most commonly a single summary tree is presented (for a given analysis), which raises the question of how good a summary tree the 19 methods produce.
% This question is related to our investigation of the accuracy of the posterior distribution on splits (via RMSD).
We use majority-rule consensus trees \citep{margush1981consensusn}, where a split appears in the consensus tree only if it appears in tree topologies whose posterior probabilities sum to at least 0.5.
Thus for two approximate posteriors to produce the same summary tree, they must only agree on whether a split probability is above or below this threshold, meaning this is a less sensitive measure of how good an approximate posterior is than RMSD or KL.
In Figure~\ref{fig:consensus_trees}, we show consensus trees for methods representing good approximations (RMSD $<$ 0.01), acceptable approximations (0.01 $\leq$ RMSD $<$ 0.05), and poor approximations (RMSD $\geq$ 0.05) for DS5 for a single run of each method.
In this run, every good approximate posterior and most (0.59\%) acceptable approximate posteriors produced a consensus tree identical to the golden run consensus tree.
A small portion (0.25\%) of poor approximate posteriors also produced identical consensus trees.

\section*{Discussion}
In this paper, we present the most comprehensive benchmark to date of methods for computing marginal likelihoods of phylogenetic tree topologies.
A number of estimators we benchmark are well known to the phylogenetics community, namely power posterior methods (\EG\ \GSS) and the HM.
We also include estimators that have been used less frequently in phylogenetics and are mainly more recent proposals: CPO, NS, and the \SHM.
Three estimators, \BS, \LPPD, and NMC, to the best of our knowledge, have not previously been used in phylogenetics.
Variational approaches have been proposed for models of heterogeneous stationary frequencies \citep{dang2018stochastic}, otherwise intractable phylogenetic models \citep{jojic2004efficient,wexler2007variational,cohn2010mean}, and to fit approximations to distributions on trees \citep{zhang2018sbn}, but to our knowledge this is the first application of the ELBO to phylogenetic model comparison.
One goal of this paper is to find methods that could work well with MCMC-free tree exploration approaches like PT, which requires evaluating the marginal likelihoods of hundreds or thousands of topologies.
Aside from the ELBO, none of the above methods are fast enough to be suitable for this purpose.
To this end, we develop the Laplus approximations and importance sampling methods based on Laplus and variational approximations.  We also consider simply using the ML and the MAP.

\subsection*{Choosing a method to use in practical scenarios}
As expected, methods differ drastically in runtime in proportion to the required Monte Carlo sampling effort.
The fastest methods took less than one second per topology on all datasets analyzed, while the slowest took over 10,000.
Perhaps surprisingly, there is no general tradeoff between speed and accuracy; while the slowest methods are among the most accurate, there are fast methods that are as good.
We break the methods down into four categories: slow, moderately slow, fast, and ultrafast, and will now reviewing the methods by category---from slow and well-known to fast and novel---highlighting the best performers and their use cases.

At the slow end of the spectrum, we find that the tried-and-true power posterior methods perform quite well, with \GSS\ providing the best (and most precise) estimates of all 19 methods.
The boost in performance compared from \GSS\ relative to the other power posterior methods comes at the cost of a marginal increase in computation time due to the estimation and multiple evaluation of the reference distribution.
The approximations produced by \PS, \PStwo, and \SStone\ are all acceptable (i.e. RMSD $< 0.05$), with most of approximations falling into the good category (i.e. RMSD $<0.01$), and are similar in terms of speed, accuracy, and precision.
The power posterior methods remain the best general-purpose tools for phylogenetic model comparisons, though they are certainly too slow to explore the tree space produced by PT.

In the middle of the speed axis, we find that \BS\ is the most promising method, with performance that is on par with \PS, \PStwo, and \SS.
As \BS\ requires an order of magnitude less time than these power posterior-based methods, if it is extended to incorporate sampling trees (perhaps following \citet{baele2015genealogical}) it could become a valuable general-purpose model selection tool.
The other estimators in this category span from poor to acceptable.
The HM is a very bad estimator of the marginal likelihood, though the related \SHM\ produces posteriors that are acceptable.
Two other methods similar in spirit to the HM, CPO (a harmonic sitewise approach) and \LPPD\ (a sitewise arithmetic approach), both perform much better than the HM or the \SHM.
NS would appear to be an unwise choice for estimating the marginal likelihoods of topologies, as it produces poor approximate posteriors.
We note that this is a somewhat different application of NS than the recent work by \citet{maturana2018nested}, who report better results of using NS when averaging over (ultrametric) trees.

\GLIS\ is the best fast method, and one of the best among the 19.
With 10,000 samples, it produces estimates of the marginal likelihood on par with \GSS\ while working three orders of magnitude more quickly.
\VBIS\ produces marginal likelihoods almost as good but is somewhat slower.
The ELBO, while faster than either \GLIS\ or \VBIS\ (which uses the variational approximation as the importance distribution) is notably worse.
It is possible that this approach suffers from getting stuck in local minima, and that multiple starting points could improve its performance, and consequently the performance of \VBIS.
The worst method in this speed category with regards to accuracy, indeed of all 19 methods, is NMC.

Among the ultrafast methods, the best candidate is \GL.
All the Laplus approximations are capable of yielding quite good estimates of the posterior distribution on trees, though they are quite variable in performance between methods, and \LL\ can produce poor approximate posteriors.
MAP and ML are faster than any of the Laplus approximations, but are not as good.
However, the success of all of these methods is truly remarkable.
Empirical posterior distributions on branch lengths are clearly not point-masses, and yet simply normalizing the unnormalized posterior at the maximum outperforms 6 of the 19 tested methods.
The success of the Laplus approximations suggests that our assumption of independence of branch lengths may not be too unreasonable, though their rather large inter-dataset variability and the improvement from importance sampling (\IE\ \GLIS) suggest that relaxing this assumption may improve performance.

\subsection*{Future directions}
We restricted ourselves here to fixed-topology inference under the simplest substitution model.
Future work should generalize beyond this simplest model to obtain a marginal likelihood across all continuous model parameters for more complex models.

Another direction for future work is to investigate the effect of modelling correlation between model parameters, including within branch lengths.
Although our preliminary results suggest that correlation between branch lengths is not strong, this assumption is not likely to hold for other parameters in more sophisticated models, such as the coalescent model in which the tree height/length is likely to be positively correlated with parameters governing population dynamics.

Another future research avenue is to find some way to reduce the inter-dataset variability of the Laplus approximations.
While this class of methods does very well on some datasets, in others there is a subset of topologies that present difficulties, possibly due to short branches with odd posterior distributions.
The problems of identifying these branches and what to do with them remain open, but solving them may greatly improve the performance of the Laplus approximation.

For fixed topology models, our results suggest bridge sampling is an accurate estimator that does not require as much compute time as the power posterior-based methods.
To apply this method more broadly to the phylogenetic field we must develop novel bridge sampling proposal distributions, perhaps modeling correlation between parameters other than branch lengths, and more importantly proposals that sample a variety of tree topologies.
However there has been some work on developing approximations of the posterior probability of trees \citep{hohna2011guided,larget2013estimation,zhang2018sbn}, notably within the \GSS\ framework \citep{baele2015genealogical}.

Another avenue for research would be to develop a diagnostic to determine an appropriate number of power posteriors that is required to accurately estimate marginal likelihoods.
Preliminary analyses have shown that the estimates calculated from 100 power posteriors were similar to estimates using 50 steps, it is however possible that fewer steps would be sufficient.

Perhaps more enticing, though, is the prospect of integrating one of the fast or ultrafast methods with PT.
PT currently uses ML---the fastest method of the 19---because speed is important, but \GL\ is comparable in speed while producing much better marginal likelihood estimates, so its inclusion in PT is worth investigating.
For the added time cost of drawing samples and calculating additional likelihoods, \GLIS\ achieves an even more impressive estimate of the marginal likelihood than \GL.
However, given that PT explores far more trees than it eventually stores, this added time cost is almost certainly prohibitive, unless the number of importance samples can be drastically reduced.
Nonetheless, once PT has found a set of high-likelihood trees, it seems prudent to use \GLIS\ on this set to produce the final approximate posterior.

\section*{Acknowledgements}
This research funded by National Science Foundation grant CISE-1561334, CISE-1564137 and National Institutes of Health grant U54-GM111274.
The research of FAM was supported in part by a Faculty Scholar grant from the Howard Hughes Medical Institute and the Simons Foundation.
Computational facilities were provided to MF by the UTS eResearch High Performance Computer Cluster.

\bibliographystyle{plainnat}
\bibliography{main}

\begin{thebibliography}{51}
\providecommand{\natexlab}[1]{#1}
\providecommand{\url}[1]{\texttt{#1}}
\expandafter\ifx\csname urlstyle\endcsname\relax
  \providecommand{\doi}[1]{doi: #1}\else
  \providecommand{\doi}{doi: \begingroup \urlstyle{rm}\Url}\fi

\bibitem[Aberer et~al.(2015)Aberer, Stamatakis, and
  Ronquist]{aberer2015efficient}
Andre~J Aberer, Alexandros Stamatakis, and Fredrik Ronquist.
\newblock An efficient independence sampler for updating branches in {Bayesian}
  {Markov} chain {Monte} {Carlo} sampling of phylogenetic trees.
\newblock \emph{Systematic Biology}, 65\penalty0 (1):\penalty0 161--176, 2015.

\bibitem[Baele et~al.(2012)Baele, Lemey, Bedford, Rambaut, Suchard, and
  Alekseyenko]{baele2012improving}
Guy Baele, Philippe Lemey, Trevor Bedford, Andrew Rambaut, Marc~A Suchard, and
  Alexander~V Alekseyenko.
\newblock Improving the accuracy of demographic and molecular clock model
  comparison while accommodating phylogenetic uncertainty.
\newblock \emph{Molecular Biology and Evolution}, 29\penalty0 (9):\penalty0
  2157--2167, 2012.

\bibitem[Baele et~al.(2015)Baele, Lemey, and Suchard]{baele2015genealogical}
Guy Baele, Philippe Lemey, and Marc~A Suchard.
\newblock Genealogical working distributions for {Bayesian} model testing with
  phylogenetic uncertainty.
\newblock \emph{Systematic Biology}, 65\penalty0 (2):\penalty0 250--264, 2015.

\bibitem[Cohn et~al.(2010)Cohn, El-Hay, Friedman, and Kupferman]{cohn2010mean}
Ido Cohn, Tal El-Hay, Nir Friedman, and Raz Kupferman.
\newblock Mean field variational approximation for continuous-time {B}ayesian
  networks.
\newblock \emph{Journal of Machine Learning Research}, 11\penalty0
  (Oct):\penalty0 2745--2783, 2010.

\bibitem[Dang and Kishino(2018)]{dang2018stochastic}
Tung Dang and Hirohisa Kishino.
\newblock Stochastic variational inference for {B}ayesian phylogenetics: A case
  of {CAT} model.
\newblock \emph{bioRxiv}, page 358747, 2018.

\bibitem[Drummond et~al.(2012)Drummond, Suchard, Xie, and
  Rambaut]{drummond2012bayesian}
Alexei~J Drummond, Marc~A Suchard, Dong Xie, and Andrew Rambaut.
\newblock Bayesian phylogenetics with {BEAUti} and the {BEAST} 1.7.
\newblock \emph{Molecular Biology and Evolution}, 29\penalty0 (8):\penalty0
  1969--1973, 2012.

\bibitem[Fan et~al.(2010)Fan, Wu, Chen, Kuo, and Lewis]{fan2010choosing}
Yu~Fan, Rui Wu, Ming-Hui Chen, Lynn Kuo, and Paul~O Lewis.
\newblock Choosing among partition models in {Bayesian} phylogenetics.
\newblock \emph{Molecular Biology and Evolution}, 28\penalty0 (1):\penalty0
  523--532, 2010.

\bibitem[Fourment and Holmes(2014)]{fourment2014novel}
Mathieu Fourment and Edward~C Holmes.
\newblock Novel non-parametric models to estimate evolutionary rates and
  divergence times from heterochronous sequence data.
\newblock \emph{BMC Evolutionary Biology}, 14\penalty0 (1):\penalty0 163, 2014.

\bibitem[Friel and Pettitt(2008)]{friel2008marginal}
Nial Friel and Anthony~N Pettitt.
\newblock Marginal likelihood estimation via power posteriors.
\newblock \emph{Journal of the Royal Statistical Society: Series B (Statistical
  Methodology)}, 70\penalty0 (3):\penalty0 589--607, 2008.

\bibitem[Friel et~al.(2014)Friel, Hurn, and Wyse]{friel2014improving}
Nial Friel, Merrilee Hurn, and Jason Wyse.
\newblock Improving power posterior estimation of statistical evidence.
\newblock \emph{Statistics and Computing}, 24\penalty0 (5):\penalty0 709--723,
  2014.

\bibitem[Gelman and Meng(1998)]{gelman1998simulating}
Andrew Gelman and Xiao-Li Meng.
\newblock Simulating normalizing constants: from importance sampling to bridge
  sampling to path sampling.
\newblock \emph{Statistical Science}, pages 163--185, 1998.

\bibitem[Gelman et~al.(1996)Gelman, Meng, and Stern]{gelman1996posterior}
Andrew Gelman, Xiao-Li Meng, and Hal Stern.
\newblock Posterior predictive assessment of model fitness via realized
  discrepancies.
\newblock \emph{Statistica Sinica}, pages 733--760, 1996.

\bibitem[Gronau et~al.(2017)Gronau, Sarafoglou, Matzke, Ly, Boehm, Marsman,
  Leslie, Forster, Wagenmakers, and Steingroever]{gronau2017tutorial}
Quentin~F Gronau, Alexandra Sarafoglou, Dora Matzke, Alexander Ly, Udo Boehm,
  Maarten Marsman, David~S Leslie, Jonathan~J Forster, Eric-Jan Wagenmakers,
  and Helen Steingroever.
\newblock A tutorial on bridge sampling.
\newblock \emph{Journal of Mathematical Psychology}, 81:\penalty0 80--97, 2017.

\bibitem[Guindon(2010)]{guindon2010bayesian}
St{\'e}phane Guindon.
\newblock Bayesian estimation of divergence times from large sequence
  alignments.
\newblock \emph{Molecular Biology and Evolution}, 27\penalty0 (8):\penalty0
  1768--1781, 2010.

\bibitem[Hammersley and Handscomb(1964)]{hammersley1964general}
John~Michael Hammersley and David~Christopher Handscomb.
\newblock General principles of the {Monte} {Carlo} method.
\newblock In \emph{Monte {Carlo} Methods}, pages 50--75. Springer, 1964.

\bibitem[Hans et~al.(2007)Hans, Dobra, and West]{hans2007shotgun}
Chris Hans, Adrian Dobra, and Mike West.
\newblock Shotgun stochastic search for ``large p'' regression.
\newblock \emph{Journal of the American Statistical Association}, 102\penalty0
  (478):\penalty0 507--516, 2007.

\bibitem[H{\"o}hna and Drummond(2011)]{hohna2011guided}
Sebastian H{\"o}hna and Alexei~J Drummond.
\newblock Guided tree topology proposals for {Bayesian} phylogenetic inference.
\newblock \emph{Systematic Biology}, 61\penalty0 (1):\penalty0 1--11, 2011.

\bibitem[H{\"o}hna et~al.(2008)H{\"o}hna, Defoin-Platel, and
  Drummond]{hohna2008clock}
Sebastian H{\"o}hna, Michael Defoin-Platel, and Alexei~J Drummond.
\newblock Clock-constrained tree proposal operators in {Bayesian} phylogenetic
  inference.
\newblock In \emph{8th IEEE International Conference on BioInformatics and
  BioEngineering (BIBE 2008)}, pages 1--7, Athens, Greece, 2008.

\bibitem[Jojic et~al.(2004)Jojic, Jojic, Meek, Geiger, Siepel, Haussler, and
  Heckerman]{jojic2004efficient}
Vladimir Jojic, Nebojsa Jojic, Chris Meek, Dan Geiger, Adam Siepel, David
  Haussler, and David Heckerman.
\newblock Efficient approximations for learning phylogenetic {HMM} models from
  data.
\newblock \emph{Bioinformatics}, 20\penalty0 (suppl\_1):\penalty0 i161--i168,
  2004.

\bibitem[Jukes and Cantor(1969)]{jukes1969evolution}
T.H Jukes and C.R. Cantor.
\newblock Evolution of protein molecules.
\newblock In \emph{Mammalian Protein Metabolism}, pages 21--32. Academic Press,
  New York, 1969.

\bibitem[Kass and Raftery(1995)]{kass1995bayes}
Robert~E Kass and Adrian~E Raftery.
\newblock Bayes factors.
\newblock \emph{Journal of the American Statistical Association}, 90\penalty0
  (430):\penalty0 773--795, 1995.

\bibitem[Kucukelbir et~al.(2015)Kucukelbir, Ranganath, Gelman, and
  Blei]{kucukelbir2015automatic}
Alp Kucukelbir, Rajesh Ranganath, Andrew Gelman, and David Blei.
\newblock Automatic variational inference in {Stan}.
\newblock In \emph{Advances in Neural Information Processing Systems}, pages
  568--576, 2015.

\bibitem[Lakner et~al.(2008)Lakner, Van Der~Mark, Huelsenbeck, Larget, and
  Ronquist]{lakner2008efficiency}
Clemens Lakner, Paul Van Der~Mark, John~P Huelsenbeck, Bret Larget, and Fredrik
  Ronquist.
\newblock Efficiency of {Markov} chain {Monte} {Carlo} tree proposals in
  {Bayesian} phylogenetics.
\newblock \emph{Systematic Biology}, 57\penalty0 (1):\penalty0 86--103, 2008.

\bibitem[Larget(2013)]{larget2013estimation}
Bret Larget.
\newblock The estimation of tree posterior probabilities using conditional
  clade probability distributions.
\newblock \emph{Systematic Biology}, 62\penalty0 (4):\penalty0 501--511, 2013.

\bibitem[Lartillot and Philippe(2006)]{lartillot2006computing}
Nicolas Lartillot and Herv{\'e} Philippe.
\newblock Computing {Bayes} factors using thermodynamic integration.
\newblock \emph{Systematic Biology}, 55\penalty0 (2):\penalty0 195--207, 2006.

\bibitem[Lenkoski and Dobra(2011)]{lenkoski2011computational}
Alex Lenkoski and Adrian Dobra.
\newblock Computational aspects related to inference in {Gaussian} graphical
  models with the {G-Wishart} prior.
\newblock \emph{Journal of Computational and Graphical Statistics}, 20\penalty0
  (1):\penalty0 140--157, 2011.

\bibitem[Lewis et~al.(2013)Lewis, Xie, Chen, Fan, and Kuo]{lewis2014posterior}
Paul~O Lewis, Wangang Xie, Ming-Hui Chen, Yu~Fan, and Lynn Kuo.
\newblock Posterior predictive {Bayesian} phylogenetic model selection.
\newblock \emph{Systematic Biology}, 63\penalty0 (3):\penalty0 309--321, 2013.

\bibitem[Lewis et~al.(2016)Lewis, Chen, Kuo, Lewis, Fu{\v{c}}{\'\i}kov{\'a},
  Neupane, Wang, and Shi]{lewis2016estimating}
Paul~O Lewis, Ming-Hui Chen, Lynn Kuo, Louise~A Lewis, Karolina
  Fu{\v{c}}{\'\i}kov{\'a}, Suman Neupane, Yu-Bo Wang, and Daoyuan Shi.
\newblock Estimating {Bayesian} phylogenetic information content.
\newblock \emph{Systematic Biology}, 65\penalty0 (6):\penalty0 1009--1023,
  2016.

\bibitem[Margush and McMorris(1981)]{margush1981consensusn}
Timothy Margush and Fred~R McMorris.
\newblock Consensusn-trees.
\newblock \emph{Bulletin of Mathematical Biology}, 43\penalty0 (2):\penalty0
  239--244, 1981.

\bibitem[Maturana~Russel et~al.(2018)Maturana~Russel, Brewer, Klaere, and
  Bouckaert]{maturana2018nested}
Patricio Maturana~Russel, Brendon~J Brewer, Steffen Klaere, and Remco~R
  Bouckaert.
\newblock Model selection and parameter inference in phylogenetics using nested
  sampling.
\newblock \emph{Systematic Biology}, June 2018.
\newblock \doi{10.1093/sysbio/syy050}.

\bibitem[Meng and Wong(1996)]{meng1996simulating}
Xiao-Li Meng and Wing~Hung Wong.
\newblock Simulating ratios of normalizing constants via a simple identity: a
  theoretical exploration.
\newblock \emph{Statistica Sinica}, pages 831--860, 1996.

\bibitem[Moler and Van~Loan(1978)]{Moler1978-gi}
C~Moler and C~Van~Loan.
\newblock Nineteen dubious ways to compute the exponential of a matrix.
\newblock \emph{SIAM Rev.}, 20\penalty0 (4):\penalty0 801--836, October 1978.

\bibitem[Moler and Van~Loan(2003)]{Moler2003-au}
C~Moler and C~Van~Loan.
\newblock Nineteen dubious ways to compute the exponential of a matrix,
  {Twenty-Five} years later.
\newblock \emph{SIAM Rev.}, 45\penalty0 (1):\penalty0 3--49, 2003.

\bibitem[Newton and Raftery(1994)]{newton1994approximate}
Michael~A Newton and Adrian~E Raftery.
\newblock Approximate {Bayesian} inference with the weighted likelihood
  bootstrap.
\newblock \emph{Journal of the Royal Statistical Society. Series B
  (Methodological)}, pages 3--48, 1994.

\bibitem[Ogata(1989)]{ogata1989monte}
Yosihiko Ogata.
\newblock A {Monte} {Carlo} method for high dimensional integration.
\newblock \emph{Numerische Mathematik}, 55\penalty0 (2):\penalty0 137--157,
  1989.

\bibitem[Overstall and Forster(2010)]{overstall2010default}
Antony~M Overstall and Jonathan~J Forster.
\newblock Default {Bayesian} model determination methods for generalised linear
  mixed models.
\newblock \emph{Computational Statistics \& Data Analysis}, 54\penalty0
  (12):\penalty0 3269--3288, 2010.

\bibitem[Owen(2013)]{mcbook}
Art~B. Owen.
\newblock \emph{Monte Carlo Theory, Methods and Examples}.
\newblock 2013.
\newblock \url{https://statweb.stanford.edu/~owen/mc/}.

\bibitem[Raftery and Banfield(1991)]{raftery1991stopping}
Adrian~E Raftery and Jeffrey~D Banfield.
\newblock Stopping the {Gibbs} sampler, the use of morphology, and other issues
  in spatial statistics ({Bayesian} image restoration, with two applications in
  spatial statistics)--(discussion).
\newblock \emph{Annals of the Institute of Statistical Mathematics},
  43\penalty0 (1):\penalty0 p32--43, 1991.

\bibitem[Ranganath et~al.(2014)Ranganath, Gerrish, and
  Blei]{ranganath2014black}
Rajesh Ranganath, Sean Gerrish, and David Blei.
\newblock Black box variational inference.
\newblock In \emph{Artificial Intelligence and Statistics}, pages 814--822,
  2014.

\bibitem[Reis and Yang(2011)]{reis2011approximate}
Mario~dos Reis and Ziheng Yang.
\newblock Approximate likelihood calculation on a phylogeny for {Bayesian}
  estimation of divergence times.
\newblock \emph{Molecular Biology and Evolution}, 28\penalty0 (7):\penalty0
  2161--2172, 2011.

\bibitem[Ronquist et~al.(2012)Ronquist, Teslenko, Van Der~Mark, Ayres, Darling,
  H{\"o}hna, Larget, Liu, Suchard, and Huelsenbeck]{ronquist2012mrbayes}
Fredrik Ronquist, Maxim Teslenko, Paul Van Der~Mark, Daniel~L Ayres, Aaron
  Darling, Sebastian H{\"o}hna, Bret Larget, Liang Liu, Marc~A Suchard, and
  John~P Huelsenbeck.
\newblock {MrBayes} 3.2: efficient {Bayesian} phylogenetic inference and model
  choice across a large model space.
\newblock \emph{Systematic Biology}, 61\penalty0 (3):\penalty0 539--542, 2012.

\bibitem[Skilling(2004)]{skilling2004nested}
John Skilling.
\newblock Nested sampling.
\newblock In \emph{AIP Conference Proceedings}, volume 735, pages 395--405.
  AIP, 2004.

\bibitem[Skilling et~al.(2006)]{skilling2006nested}
John Skilling et~al.
\newblock Nested sampling for general {Bayesian} computation.
\newblock \emph{Bayesian Analysis}, 1\penalty0 (4):\penalty0 833--859, 2006.

\bibitem[Thorne et~al.(1998)Thorne, Kishino, and Painter]{thorne1998estimating}
Jeffrey~L Thorne, Hirohisa Kishino, and Ian~S Painter.
\newblock Estimating the rate of evolution of the rate of molecular evolution.
\newblock \emph{Molecular Biology and Evolution}, 15\penalty0 (12):\penalty0
  1647--1657, 1998.

\bibitem[Tierney and Kadane(1986)]{tierney1986accurate}
Luke Tierney and Joseph~B Kadane.
\newblock Accurate approximations for posterior moments and marginal densities.
\newblock \emph{Journal of the American Statistical Association}, 81\penalty0
  (393):\penalty0 82--86, 1986.

\bibitem[Vehtari et~al.(2017)Vehtari, Gelman, and Gabry]{vehtari2017practical}
Aki Vehtari, Andrew Gelman, and Jonah Gabry.
\newblock Practical {Bayesian} model evaluation using leave-one-out
  cross-validation and {WAIC}.
\newblock \emph{Statistics and Computing}, 27\penalty0 (5):\penalty0
  1413--1432, 2017.

\bibitem[Wexler and Geiger(2007)]{wexler2007variational}
Ydo Wexler and Dan Geiger.
\newblock Variational upper bounds for probabilistic phylogenetic models.
\newblock In \emph{Annual International Conference on Research in Computational
  Molecular Biology}, pages 226--237. Springer, 2007.

\bibitem[Whidden and Matsen~IV(2015)]{whidden2015quantifying}
Chris Whidden and Frederick~A Matsen~IV.
\newblock Quantifying {MCMC} exploration of phylogenetic tree space.
\newblock \emph{Systematic Biology}, 64\penalty0 (3):\penalty0 472--491, 2015.

\bibitem[Whidden et~al.(2018)Whidden, Claywell, Fisher, Magee, Fourment, and
  Matsen~IV]{pt}
Chris Whidden, Brian Claywell, Thayer Fisher, Andy Magee, Mathieu Fourment, and
  Frederick~A Matsen~IV.
\newblock Systematic exploration of the high likelihood density set of
  phylogenetic trees.
\newblock 2018.

\bibitem[Xie et~al.(2010)Xie, Lewis, Fan, Kuo, and Chen]{xie2010improving}
Wangang Xie, Paul~O Lewis, Yu~Fan, Lynn Kuo, and Ming-Hui Chen.
\newblock Improving marginal likelihood estimation for {Bayesian} phylogenetic
  model selection.
\newblock \emph{Systematic Biology}, 60\penalty0 (2):\penalty0 150--160, 2010.

\bibitem[{Zhang} and {Matsen}(2018)]{zhang2018sbn}
C.~{Zhang} and F.~A. {Matsen}, IV.
\newblock {Generalizing tree probability estimation via {Bayesian} networks}.
\newblock \emph{ArXiv e-prints}, May 2018.

\end{thebibliography}

\beginsupplement
\clearpage

\section{Methods}

In the Bayesian framework, the marginal likelihood or evidence of data $D$ conditioned on model $\tau$ with associated parameters $\boldsymbol{\theta} = (\theta_1, \theta_2, \dots, \theta_N)$ is
\[
p(D \mid \tau) = \int p(D \mid \boldsymbol{\theta}, \tau) p(\boldsymbol{\theta} \mid \tau) d\boldsymbol{\theta},
\]
where $p(D \mid \boldsymbol{\theta}, \tau)$ is the probability of the data given parameters $\boldsymbol{\theta}$, $p(\boldsymbol{\theta} \mid \tau)$ is the prior on $\boldsymbol{\theta}$, and the integral is of dimension N.

Dependence on model $\tau$ is suppressed in the rest of the document to simplify notation.

\subsection{Laplace method}

\subsubsection{Classical Laplace}
The Laplace method \citep{tierney1986accurate} approximates the marginal likelihood by approximating the posterior distribution using a multivariate normal distribution with mean equal to the maximum a posteriori estimates $\boldsymbol{\tilde{\theta}}$, and covariance $\tilde{\Sigma}=(-H)^{-1}$ where $H$ is the Hessian matrix of second derivatives of $\log(p(D \mid \boldsymbol{\theta})p(\boldsymbol{\theta}))$.
Specifically, let us define $l(\boldsymbol{\theta}) = \log(p(D \mid \boldsymbol{\theta})p(\boldsymbol{\theta}))$ and Taylor-expand $l(\boldsymbol{\theta})$ around $\boldsymbol{\tilde{\theta}}$.
Exponentiating this quadratic approximation leads to a normal distribution  with $\boldsymbol{\tilde{\mu}} = \boldsymbol{\tilde{\theta}}$ and $\tilde{\Sigma} = -H^{-1}$.
Integrating the normal distribution yields the Laplace marginal likelihood estimator

\[
\hat{p}_{L}(D) \approx (2 \pi)^{d/2} \det(\tilde{\Sigma})^{1/2} p(D \mid \boldsymbol{\tilde{\theta}}) p(\boldsymbol{\tilde{\theta}}),
\]
where $\det(\tilde{\Sigma})$ is the determinant of the covariance matrix.

Unfortunately, the above normal approximation is not always accurate in practice. In our specific phylogenetic setting, the positivity of branch lengths creates problems for the normal approximation.
It is however possible to improve the normal approximation of the posterior and the Laplace method if we transform each variable $\theta_i$ using a one-to-one twice differentiable function $g$ such as $\theta_i = g(z_i)$ and $z_i = g^{-1}(\theta_i)$.
Applying the chain rule, the Hessian of the posterior for the transformed parameters is
\[
H_{i,j}^z = \frac{\partial^2 l}{\partial z_i \partial z_j} =
\begin{cases}
\frac{\partial l}{\partial \theta_i} \frac{\partial^2 \theta_i}{\partial z_i^2} + H_{ii} \left(\frac{\partial \theta_i}{\partial z_i}\right)^2 & \text{for } i = j, \\
H_{ij} \frac{\partial \theta_i}{\partial z_i} \frac{\partial \theta_j}{\partial z_j} & \text{otherwise}.
\end{cases}
\]

The transformation requires an adjustment to account for the distortion of the distribution hence insuring that the distribution integrates to 1.
Therefore, given $\boldsymbol{z} \sim \mathcal{N}(\boldsymbol{\mu} ,\boldsymbol{\Sigma })$ the density of $\boldsymbol\theta$ is
\[
p(\boldsymbol{\theta}) = \mathcal{N}(g^{-1}(\boldsymbol\theta) \mid \boldsymbol{\mu} ,\boldsymbol{\Sigma })\, | \det J_{g^{-1}}(\boldsymbol\theta) |,
\]
where $| \det J_{g^{-1}}(\boldsymbol\theta) |$ is the absolute value of the determinant of the Jacobian matrix evaluated at $\boldsymbol\theta$.
However, we find in practice that some branch length posteriors are monotonically decreasing functions with modes at 0, and thus the transformation approach is not sufficient to make the normal approximation accurate.

\subsubsection{The Laplus approximations}

However, while some transformations may work well for a branch or subset of branches, we find in practice that there is no one transformation that works well for all branches on a tree.
As an alternative we use a family of approximations inspired by the Laplace that we call the Laplus approximations (in recognition of the fact that they are like the Laplace but designed for parameters on $\mathbb{R}^+$).
We share with the Laplace approximation the assumption that the posterior is concentrated around the mode, $\boldsymbol{\tilde{\theta}}$.
Unlike the Laplace approximation, we assume that branch lengths are mutually independent, such that we can make the approximation
\[
p(\boldsymbol{\theta} \mid \tau, D) \approx \prod_i q(\theta_i;\boldsymbol{\phi}_i)
\]
Here $q$ is a parametric distribution with known normalizing constant (such as the gamma distribution) that we will use to approximate the posterior distributions for each branch.
For a given branch, $\boldsymbol{\phi}_i$ are the parameters of $q$ that approximate the marginal posterior of that branch.
Let $C$ be a constant such that
\[
p(\boldsymbol{\theta} \mid \tau, D) = C \times p(D \mid \tau, \boldsymbol{\theta}) p(\boldsymbol{\theta})
\]
That is, $C$ is the inverse of the marginal likelihood that we seek to estimate, and using our approximation above,
\[
C = \frac{p(\boldsymbol{\theta} \mid \tau, D)}{p(D \mid \tau, \boldsymbol{\theta}) p(\boldsymbol{\theta})} \approx \frac{\prod_i q(\theta_i;\boldsymbol{\phi}_i)}{p(D \mid \tau, \boldsymbol{\theta}) p(\boldsymbol{\theta})}
\]
Finally, by applying this equation at the posterior mode, our resulting estimate of the marginal likelihood is
\[
\hat{p}_{\text{Laplus}}(D) = \hat{C}^{-1} = \frac{p(D|\boldsymbol{\tilde{\theta}})p(\boldsymbol{\tilde{\theta}})}{\prod_i q(\tilde{\theta}_i ; \boldsymbol{\phi}_{i})}
\]

The general procedure for the Laplus approximations is similar regardless of parametric distributional family assumption $q$.
Our goal is to take the joint MAP estimates of the branch lengths $\boldsymbol{\tilde{\theta}}$ and the vector of second derivatives of the log-posterior $(\frac{\partial^2l}{\partial \theta_1^2}, \frac{\partial^2l}{\partial \theta_2^2}, \dots, \frac{\partial^2l}{\partial \theta_n^2})$ and find the parameters of our approximating distributions for each branch, $\boldsymbol{\phi}_i$, by matching modes and second derivatives of the approximating and posterior distributions of branch lengths.
The complete procedure is written here algorithmically.

\renewcommand{\labelenumii}{(\roman{enumii})}
\begin{enumerate}
  \item Find the (joint) MAP branch lengths, $\boldsymbol{\tilde{\theta}} = (\tilde{\theta}_1,\tilde{\theta}_2, \dots ,\tilde{\theta}_n)$
  \item for $i$ in $1:n$
    \begin{enumerate}
      \item Compute $\frac{\partial^2l}{\partial \theta_i^2}$, the second derivative of the log unnormalized posterior with respect to the $i^{\textnormal{th}}$ branch
      \item Find parameters of $\boldsymbol{\phi}_{i} $ by solving
\begin{align*}
\frac{d^2}{dx^2}\log(q(x; \boldsymbol{\phi}_{i})) &= \frac{\partial^2l}{\partial \theta_i^2}\Big|_{\theta_i = \tilde{\theta}_i} \\
\text{mode}(q(x; \boldsymbol{\phi}_{i})) &= \tilde{\theta}_i
	\end{align*}
      \item Catch exceptions
    \end{enumerate}
  \item Compute the marginal likelihood as $\hat{p}_{\text{Laplus}}(D) = \frac{p(D|\boldsymbol{\tilde{\theta}})p(\boldsymbol{\tilde{\theta}})}{\prod_i q(\tilde{\theta}_i ; \boldsymbol{\phi}_{i})}$
\end{enumerate}

Exceptions occur when elements of $\boldsymbol{\phi}_{i}$ are outside of the domain of support, when $H_{ii}$ is nonnegative (so the posterior has a mode at 0), or when elements of $\boldsymbol{\phi}_{i}$ are otherwise suspect (such as producing particularly high-variance distributions with very short branches).
Exceptions and their handling depend on the distributional assumption, and so we describe exception handling in the section for each distribution individually.
We consider three choices for $q$, the gamma distribution, the \Betaprime\ distribution, and the lognormal distribution.
Since the Laplus method is not derived through a Taylor expansion of the unnormalised posterior, it is not subject to some of the assumptions required by Laplace's method.
Although both methods require the function to be twice differentiable, Laplace's method assumes that the global maxima $\tilde{\boldsymbol{\theta}}$ is not at the boundary of the interval of integration so that the first derivatives vanishes at $\tilde{\boldsymbol{\theta}}$.
Zero-length branches have typically non-zero (i.e. negative) first derivatives and positive second derivatives making the Laplus method attractive.
And while it is obvious that there must be some dependence between branch lengths, we find in practice that the posterior correlations between branch lengths are often quite small.

\subsubsection{Gamma-Laplus}

Here we seek to approximate the marginal posteriors of all branch lengths with gamma distributions.
The vector $\boldsymbol{\phi}_{i} = (\alpha_i, \beta_i)$ contains the shape and rate parameters of the gamma distribution; the log probability density function of the gamma is
\[
\log(\text{Gamma}(x;\alpha,\beta)) = \alpha \log(\beta) - \log(\Gamma(\alpha)) + (\alpha - 1) \log(x) - \beta x.
\]

The first and second derivatives of the log gamma distribution with respect to $x$ are given by

\begin{align*}
  \frac{d}{d x}\log(\text{Gamma}(x;\alpha,\beta)) &= \frac{\alpha - 1}{x} - \beta,\\
  \frac{d^2}{d x^2}\log(\text{Gamma}(x;\alpha,\beta)) &= -\frac{\alpha - 1}{x^2}.\\
\end{align*}

We make use of the second derivative of the log-posterior at the mode, $H_{ii} = \frac{\partial^2l}{\partial \theta_i^2}\Big|_{\theta_i = \tilde{\theta}_i}$ to estimate $\hat{\alpha}_i$ using the second derivative of the log of the gamma distribution.
Then we solve for $\hat{\beta}_i$ using the analytic formula for gamma mode: $\tilde{\theta}_i = \frac{\alpha_i - 1}{\beta_i}$.
\begin{align*}
  H_{ii} &= -\frac{\hat{\alpha}_i - 1}{\tilde{\theta}_i^2}\\
  \hat{\alpha}_i &= 1 - \tilde{\theta}_i^2 H_{ii}\\
  \hat{\beta}_i &= \frac{\hat{\alpha}_i - 1}{\tilde{\theta}_i}
  = \frac{-\tilde{\theta}_i^2 H_{ii}}{\tilde{\theta}_i} = -\tilde{\theta}_i H_{ii}
\end{align*}

We note two exceptions to handle with the \GL\ approach.
The first case are branches with a mode at 0, which have posteriors that are monotonically decreasing.
The second case are branches that are short with oddly large variances.
We detect branches of the first type by checking whether $\tilde{\theta}_i < \epsilon_1$ or $H_{ii} >= 0$.
These branches are handled by fixing $\hat{\alpha_i} = 1$ (to ensure that the approximation is monotonically decreasing) and fitting $\hat{\beta_i}$ directly using the log-posterior calculated at $N$ points spaced evenly (on the log-scale) between $\tilde{\theta}_i$ and 0.5.
We detect branches of the second type by checking whether $\tilde{\theta}_i < \epsilon_2$ and $\frac{\alpha_i}{\beta_i^2} > 0.1$.
These branches are handled by fitting $\alpha_i, \beta_i$ to $N$ points spaced evenly (on the log-scale) between $\tilde{\theta}_i$ and 0.5, while constraining $\tilde{\theta}_i = \frac{\hat{\alpha}_i - 1}{\hat{\beta}_i}$ (such that the mode of the approximation to be the mode of the posterior).
We use $N = 10$, $\epsilon_1 = 10^{-6}$, and $\epsilon_2 = 10^{-4}$.

\subsubsection{\Betaprime-Laplus}

Here we seek to approximate the marginal posteriors of all branch lengths as beta$'$ distributions.
In this case, the vector $\boldsymbol{\phi}_{i} = (\alpha_i, \beta_i)$ concatenates the shape parameters of the beta$'$  distribution with log probability density function is
\[
\log(\text{Beta}'(x;\alpha,\beta)) = -\log(B(\alpha,\beta)) + (\alpha - 1) \log(x) - (\alpha+\beta)\log(x+1),
\]

where $B$ is the beta function.

The first and second derivatives of the log beta$'$ distribution with respect to $x$ are given by

\begin{align*}
  \frac{d}{d x}\log(\text{Beta}'(x;\alpha,\beta)) &= \frac{\alpha - 1}{x} - \frac{\alpha + \beta}{x + 1},\\
  \frac{d^2}{d x^2}\log(\text{Beta}'(x;\alpha,\beta)) &= -\frac{\alpha - 1}{x^2} + \frac{\alpha + \beta}{(x + 1)^2}.\\
\end{align*}

When $\alpha \leq 1$, the beta$'$ distribution collapses to a monotonically decreasing distribution.
When $\alpha = 1$,
\begin{align*}
  \log(\text{Beta}'(x;1,\beta_i)) &= -\log(B(1,\beta_i)) + (1 - 1) \log(x) - (1+\beta_i)\log(x+1),\\
  \log(\text{Beta}'(x;1,\beta_i)) &= -\log(B(1,\beta_i)) - (1+\beta_i)\log(x+1),\\
  \frac{d}{d x}\log(\text{Beta}'(x; 1,\beta_i)) &=  -\frac{1 + \beta_i}{x + 1}.
\end{align*}

We make use of the second derivative at the mode, $H_{ii} = \frac{\partial^2l}{\partial \theta_i^2}\Big|_{\theta_i = \tilde{\theta}_i}$ to estimate $\hat{\beta}_i$. Then we solve for $\hat{\alpha}_i$ using the fact that $\tilde{\theta}_i = \frac{\hat{\alpha}_i - 1}{\hat{\beta}_i + 1}$.

\begin{align*}
  H_{ii} &= -\frac{\hat{\alpha}_i - 1}{\tilde{\theta}_i^2} + \frac{\hat{\alpha}_i + \hat{\beta}_i}{(\tilde{\theta}_i + 1)^2}\\
  &= -\frac{1}{\tilde{\theta}_i} \frac{\hat{\alpha}_i - 1}{\frac{\hat{\alpha}_i - 1}{\hat{\beta}_i + 1}} + \frac{1}{\tilde{\theta}_i + 1} \frac{\hat{\alpha}_i + \hat{\beta}_i}{\frac{\hat{\alpha}_i + \hat{\beta}_i}{\hat{\beta}_i + 1}}\\
  &= -\frac{1}{\tilde{\theta_i}} (\hat{\beta}_i + 1) + \frac{1}{\tilde{\theta}_i + 1} (\hat{\beta}_i + 1)\\
  &=  (\hat{\beta}_i + 1) \Big(\frac{1}{\tilde{\theta}_i + 1} - \frac{1}{\tilde{\theta}_i}\Big)\\
  &=  \frac{\hat{\beta}_i + 1}{\tilde{\theta}_i(\tilde{\theta}_i + 1)}\\
  \hat{\beta}_i &= -H_{ii} (\tilde{\theta}_i + 1) \tilde{\theta}_i - 1\\
  \hat{\alpha}_i &= \tilde{\theta}_i(\hat{\beta}_i + 1) + 1.
\end{align*}

We note two exceptions to handle with the BL approach.
To start, we check if $\hat{\beta}_i < 0$, which implies $H_{ii} > 0$, meaning the posterior should be monotonically decreasing.
In this case, we set $\hat{\alpha}_i = 1$ and use the equations outlined below to fit $\hat{\beta}_i$.
We then check if $\hat{\beta}_i < 2$, in which case our approximate posterior has suspiciously high variance, in which case we fit $\alpha_i, \beta_i$ to $N$ points spaced evenly (on the log-scale) between $\tilde{\theta}_i$ and 0.5, while constraining $\tilde{\theta}_i = \frac{\hat{\alpha}_i - 1}{\hat{\beta}_i + 1}$ (such that the mode of the approximation to be the mode of the posterior).

When we set $\hat{\alpha}_i = 1$ we can use the first derivative of the log-posterior, $\nabla_{i} = \frac{\partial l}{\partial \theta_i} \Big|_{\theta_i = \tilde{\theta}_i}$, to fit $\hat{\beta}_i$:
\begin{align*}
  \nabla_i &= -\frac{1 + \hat{\beta}_i}{\tilde{\theta}_i + 1},\\
  \hat{\beta_i} &= -\nabla_i (\tilde{\theta}_i + 1) - 1.
\end{align*}

\subsubsection{Lognormal-Laplus}

Here we seek to approximate the marginal posteriors of all branch lengths as lognormal distributions.
The vector $\boldsymbol{\phi}_{i} = (\mu_i, \sigma_i)$ concatenates the mean and standard deviation parameters of the lognormal  distribution with log probability density function
\[
\log(\text{Lognormal}(x;\mu_i,\sigma_i)) = -\frac{\log(2\pi)}{2} - \log(x) - \log(\sigma_i) - \frac{(\log(x) - \mu_i)^2}{2\sigma_i^2}.
\]

The first and second derivatives of the log lognormal distribution with respect to $x$ are given by

\begin{align*}
  \frac{d}{d x}\log(\text{Lognormal}(x;\mu_i,\sigma_i)) &= -\frac{1}{x} - \frac{\log(x) - \mu_i}{x\sigma_i^2},\\
  \frac{d^2}{d x^2}\log(\text{Lognormal}(x;\mu_i,\sigma_i)) &= \frac{1}{x^2} - \frac{-\log(x) + \mu_i + 1}{x^2\sigma_i^2}.\\
\end{align*}

We make use of the second derivative at the mode, $H_{ii} = \frac{\partial^2l}{\partial \theta_i^2}\Big|_{\theta_i = \tilde{\theta}_i}$, and the fact that $\tilde{\theta}_i = e^{\mu_i - \sigma_i^2}$ to estimate $\hat{\sigma}_i^2$.
Then we solve for $\hat{\mu}_i$ using the fact that $\log(\tilde{\theta}_i) = \mu_i - \sigma_i^2$.
\begin{align*}
  H_{ii} &= \frac{1}{\tilde{\theta}_i^2} - \frac{-\log(\tilde{\theta}_i) + \hat{\mu}_i + 1}{\tilde{\theta}_i^2 \hat{\sigma}_i^2}\\
  &= \frac{1}{\tilde{\theta}_i^2} - \frac{-(\hat{\mu}_i - \hat{\sigma}_i^2) + \hat{\mu}_i + 1}{\tilde{\theta}_i^2 \hat{\sigma}_i^2}\\
  &= \frac{1}{\tilde{\theta}_i^2} - \frac{1}{\tilde{\theta}_i^2 \hat{\sigma}_i^2} - \frac{\hat{\sigma}_i^2}{\tilde{\theta}_i^2 \hat{\sigma}_i^2},\\
  \hat{\sigma}_i^2 &= - \frac{1}{\tilde{\theta}_i^2 H_{ii}},\\
  \hat{\mu}_i &= \log(\tilde{\theta}_i) + \hat{\sigma}_i^2.\\
\end{align*}

We note two exceptions to handle with the \LL\ approach.
The first case are branches with a mode at 0, which have posteriors that are monotonically decreasing.
The second case are branches that are short with oddly large variances.
We nest the cases such that we first check for branches that fall in either category, checking $\tilde{\theta}_i < \epsilon_1$ or $H_{ii} >= 0$ or $\hat{\mu} > 5$ (which happens when $\hat{\sigma}$ is suspiciously large).
As there is no parameter regime in which the lognormal is monotonically decreasing, and suspiciously high-variance branches are not fit any better by a lognormal distribution than a gamma distribution, at this point we switch to approximating branches as gamma distributions and proceed with exceptions as in the \GL\ approach.

\subsection{Importance sampling}

Importance sampling uses a reference or importance distribution from which values are drawn, allowing summaries to be calculated for an unknown distribution by taking into account the importance weights (probabilities of drawing the sampled values). If $g$ is an importance distribution then
\[
\begin{aligned}
p(D) &= \int p(D \mid \boldsymbol{\theta}) p(\boldsymbol{\theta}) \text{d}\boldsymbol{\theta}\\
&= \int  \frac{p(D \mid \boldsymbol{\theta}) p(\boldsymbol{\theta})} {g(\boldsymbol{\theta})} g(\boldsymbol{\theta}) \text{d}\boldsymbol{\theta} \\
&= \mathbb{E}_g \left( \frac{p(D \mid \boldsymbol{\theta}) p(\boldsymbol{\theta})} {g(\boldsymbol{\theta})} \right).
\end{aligned}
\]

For a normalized density $g$, the estimate is given by,
\[
\hat{p}_{\text{IS}}(D) = \frac{1}{N} \sum_{i=1}^N \frac{p(D \mid \tilde{\boldsymbol{\theta}_i}) p(\tilde{\boldsymbol{\theta}_i})}{g(\tilde{\boldsymbol{\theta}_i})}, \tilde{\boldsymbol{\theta}}_i \sim g(\boldsymbol{\theta}).
\]

For an unnormalized density $q$, the self normalized importance sampling estimate \citep{mcbook} is given by

\[
\hat{p}_{\text{IS}}(D) = \frac{\sum_{i=1}^N p(D \mid \tilde{\boldsymbol{\theta}}_i) w(\tilde{\boldsymbol{\theta}}_i)}{\sum_{i=1}^N w(\tilde{\boldsymbol{\theta}}_i)}, \tilde{\boldsymbol{\theta}}_i \sim q(\boldsymbol{\theta}),
\]
where $w(\tilde{\boldsymbol{\theta}}_i)$ is the importance weight given by  $w(\tilde{\boldsymbol{\theta}}_i)=\frac{p(\tilde{\boldsymbol{\theta}}_i)}{q(\tilde{\boldsymbol{\theta}}_i)}$.

\subsection{Naive Monte Carlo}

The simplest Monte Carlo estimator of the marginal likelihood is defined as the expected value of the likelihood with respect to the prior distribution \citep{hammersley1964general,raftery1991stopping}.
The so called naive Monte Carlo (\NMC) estimator can be approximated by drawing $N$ samples ${\boldsymbol{\theta}_1, \boldsymbol{\theta}_2, \dots, \boldsymbol{\theta}_N}$ from the prior distribution and calculating the arithmetic mean of the likelihood.

\[
\hat{p}_{\text{NMC}}(D) = \frac{1}{N} \sum_{i=1}^N p(D \mid \tilde{\boldsymbol{\theta}_i}), \tilde{\boldsymbol{\theta}}_i \sim p(\boldsymbol{\theta}).
\]

Although this approach is fast and unbiased, the high-likelihood region can be distant from the high-prior region.
Most $\tilde{\boldsymbol{\theta}}_i$s will therefore be sampled from a region of the likelihood with low probability yielding high variance \citep{newton1994approximate}.

% \subsection{Arithmetic mean}
% \citep{aitkin1991posterior} proposed to sample from the posterior instead of the prior distribution and use the arithmetic mean of the likelihood.
%
% \[
% \hat{p}_{AM}(D) = \frac{1}{N} \sum_{i=1}^N p(D \mid \tilde{\theta_i}), \tilde\theta_i \sim p(\theta \mid D)
% \]
%
% While this approach is appealing since the practitioner can use samples generated by an MCMC and it avoids problems with vague priors, it was strongly criticized by the Bayesian community.
% The AM estimator overstates the fit of the model to the data by using the data twice \citep{kadane2004methods}.

\subsection{Harmonic mean}
The harmonic mean (\HM) estimator only requires  samples from the posterior generated by a single MCMC or other samplers and is therefore appealing to the user \citep{newton1994approximate}.
The harmonic mean estimator of marginal estimator is equivalent to an importance sampling estimator of $1/p(D)$ with importance distribution $p(\boldsymbol{\theta} \mid D)$:

\[
\hat{p}_{\text{HM}}(D) = \frac{1}{\frac{1}{N} \sum_{i=1}^N \frac{1}{p(D \mid \tilde{\boldsymbol{\theta}_i})}}, \tilde{\boldsymbol{\theta}}_i \sim p(\boldsymbol{\theta} \mid D).
\]

This estimator is unstable due to the possible occurrence of small likelihood values the estimator and hence this estimator has infinite variance.
Although the Law of Large Numbers guarantees that this estimator is consistent, the number of samples required to get an accurate estimate can be prohibitively high.

\subsection{Stabilized harmonic mean}
\citet{newton1994approximate} also proposed the stabilized harmonic mean (\SHM) estimator to address the instability of the \HM\ estimator.
The \SHM\ estimator is based on importance sampling scheme where the importance sampling distribution is a mixture of the prior and the posterior: $p^\star(\boldsymbol{\theta}) = \delta p(\boldsymbol{\theta}) + (1 - \delta)p(\boldsymbol{\theta} \mid D)$ where $\delta$ is small, such that

\[
\hat{p}_{\text{SHM}^*}(D) = \frac{\sum_{i=1}^n \frac{p(D \mid \tilde{\boldsymbol{\theta}}_i)}{\delta \hat{p}_{\text{SHM}^*}(D) + (1 - \delta)p(D \mid \tilde{\boldsymbol{\theta}}_i)} }{\sum_{i=1}^n \{\delta \hat{p}_{\text{SHM}^*}(D) + (1 - \delta)p(D \mid \tilde{\boldsymbol{\theta}}_i)\}^{-1}}, \tilde{\boldsymbol{\theta}}_i \sim p^\star(\boldsymbol{\theta}).
\]

Unfortunately this estimator requires simulating from both the posterior and prior.
Newton and Raftery proposed to simulate from the posterior and assume that a further $\frac{\delta n}{(1-\delta)}$ observations are drawn from the prior, all of them with their likelihoods equal to their expected value $p(D)$.
The likelihood of the imaginary samples drawn from the prior is $p(D \mid \theta_j) = \hat{p}_{SHM}$ for $j=1, \dots, \frac{\delta n}{1 - \delta}$.
Then, the approximate marginal likelihood $\hat{p}_{\text{SHM}}(D)$ satisfies the following equation:
\[
\hat{p}_{\text{SHM}}(D) = \frac{\frac{\delta n}{1 - \delta} + \sum_{i=1}^n \frac{p(D \mid \tilde{\boldsymbol{\theta}}_i)}{\delta \hat{p}_{\text{SHM}}(D) + (1 - \delta)p(D \mid \tilde{\boldsymbol{\theta}}_i)}}{\frac{\delta n}{(1-\delta)\hat{p}_{\text{SHM}}(D)} + \sum_{i=1}^n \{\delta \hat{p}_{\text{SHM}}(D) + (1 - \delta)p(D \mid \tilde{\boldsymbol{\theta}}_i)\}^{-1}}, \tilde{\boldsymbol{\theta}}_i \sim p(\boldsymbol{\theta} \mid D),
\]
which is solved by  an iterative scheme that updates an initial guess of the marginal likelihood (e.g. harmonic mean estimate) until a stopping criterion is satisfied.
In our implementation the recursion stops when the absolute change in $\log \hat{p}_{\text{SHM}}(D)$ is less than $10^{-7}$.
\citet{newton1994approximate} advocate $\delta = 0.01$ while \citet{lartillot2006computing} use $\delta = 0.1$.
In this study we used the $\hat{p}_{\text{SHM}}$ with $\delta = 0.01$.

\subsection{Bridge sampling}
Bridge sampling (\BS) was initially developed to estimate Bayes factors \citep{kass1995bayes} and was more recently adapted to approximate the marginal likelihood of a single model \citep{overstall2010default,gronau2017tutorial}.
Following a derivation by \citet{gronau2017tutorial}, the bridge sampling estimator is derived from the following identity:
\[
1 = \frac{\int p(D \mid \boldsymbol{\theta}) p(\boldsymbol{\theta}) h(\boldsymbol{\theta}) g(\boldsymbol{\theta}) d \boldsymbol{\theta}}{\int p(D \mid \boldsymbol{\theta}) p(\boldsymbol{\theta}) h(\boldsymbol{\theta}) g(\boldsymbol{\theta}) d \boldsymbol{\theta}},
\]
where $g(\boldsymbol{\theta})$ is the proposal distribution and $h(\boldsymbol{\theta})$ is the bridge function.
The bridge function ensures that the denominator in the identity is not zero.

Multiplying both sides of the above identity by $p(D)$ the bridge sampling estimator of the marginal likelihood is

\[
p_{\text{BS}}(D) = \frac{\int p(D \mid \boldsymbol{\theta}) p(\boldsymbol{\theta}) h(\boldsymbol{\theta}) g(\boldsymbol{\theta}) d \boldsymbol{\theta}}{\int h(\boldsymbol{\theta}) g(\boldsymbol{\theta}) p(\boldsymbol{\theta} \mid D) d \boldsymbol{\theta}} = \frac{\mathbb{E}_{g(\boldsymbol{\theta})} (p(D \mid \boldsymbol{\theta}) p(\boldsymbol{\theta}) h(\boldsymbol{\theta}))}{\mathbb{E}_{p(\boldsymbol{\theta} \mid D)} (h(\boldsymbol{\theta}) g(\boldsymbol{\theta}))}.
\]

The marginal likelihood is approximated using $n_1$ samples from the posterior distribution and $n_2$ samples from the proposal distribution
\[
\hat{p}_{\text{BS}}(D) = \frac{1/n_2 \sum_{i=1}^{n_2} (p(D \mid \tilde{\boldsymbol{\theta}_i}) p(\tilde{\boldsymbol{\theta}}_i) h(\tilde{\boldsymbol{\theta}}_i))}{1/n_1 \sum_{j=1}^{n_1} h(\boldsymbol{\theta}_j^*) g(\boldsymbol{\theta}_j^*)}, \tilde{\boldsymbol{\theta}}_i \sim g(\boldsymbol{\theta}), \boldsymbol{\theta}_j^* \sim p(\boldsymbol{\theta} \mid D).
\]

Several bridge functions can be used including the so called \textit{optimal bridge function} \citep{meng1996simulating}:
\[
h(\boldsymbol{\theta}) = \frac{C}{s_1 p(D \mid \boldsymbol{\theta}) p(\boldsymbol{\theta}) + s_2 p(D) g(\boldsymbol{\theta})},
\]
where $s_1 = n_1/(n_1 + n_2)$ and $s_2 = n_2/(n_1 + n_2)$ and $C$ is a constant that cancels out.

The definition of the optimal bridge function depends on the marginal likelihood itself, suggesting an iterative scheme to approximate $p(D)$ starting from an initial guess, such as the \HM\ estimate.
\citet{gronau2017tutorial} provide a detailed description of an algorithm.

\subsection{Thermodynamic integration (aka path sampling, power posterior)}
The thermodynamic integration estimator was introduced by \citet{lartillot2006computing} in the phylogenetic context, borrowing ideas from path sampling \citep{gelman1998simulating} and the physics literature where a large body of research is dedicated to the estimation of normalisation constants.
Lartillot and Philippe defined a path going from the prior to the unnormalised posterior $q$ using
\[
q_\beta = p(D \mid \boldsymbol{\theta})^\beta p(\boldsymbol{\theta})
\]
for $\beta \in [0,1]$.
The normalisation constant $Z_\beta$ of the tempered unnormalised posterior is therefore
\[
Z_\beta = \int_{\boldsymbol{\theta}} p(D \mid \boldsymbol{\theta})^\beta p(\boldsymbol{\theta}) d\boldsymbol{\theta}
\]
and the log marginal likelihood of the model follows from the path sampling identity:
\[
\log p(D) = \log Z_1 - \log Z_0
 = \int_0^1 \frac{\partial Z_\beta}{\partial \beta} d\beta
 = \int_0^1 E_{\boldsymbol{\theta} \mid D,\beta}(\log p(D \mid \boldsymbol{\theta})) d\beta .
\]

\citet{friel2008marginal}  worked on similar ideas but differ in the choice of temperature schedule and how the integral over [0,1] is approximated.
\citet{lartillot2006computing} approximate the integral using the  Simpson's rule while  \citet{friel2008marginal} applied the trapezoidal rule.
The interval $\beta \in [0,1]$ is discretized such that $0=\beta_0 < \beta_1< \dots < \beta_K = 1$ and for each $\beta_i$ samples are drawn from $p(\boldsymbol{\theta} \mid D, \beta_i)$ to estimate $E_{\boldsymbol{\theta} \mid D, \beta_i}( \log p(D \mid \boldsymbol{\theta}))$.
For example, using the trapezoidal rule the log marginal likelihood of a given model is
\[
\log \hat{p}_{\text{PS}}(D) \approx \sum_{i=1}^K (\beta_i - \beta_{i-1})\left(\frac{E_{i-1} + E_i}{2}\right),
\]
where $E_i = E_{\boldsymbol{\theta} \mid \beta_i} \log p(D \mid \boldsymbol{\theta})$ is the expectation of the log deviance at $\beta_i$.

\citet{lartillot2006computing} used equally spaced inverse temperatures between 0 and 1, while \citet{friel2008marginal} set $\beta_i = (i/K)^5$.
It is clear that other temperature schedules can be exploited such as a schedule based on the quantiles of parametric distribution \citep{xie2010improving} (see stepping stone section) and the adaptive scheme proposed by \citet{friel2014improving}.
\citet{friel2014improving} subsequently proposed a modified trapezoidal rule that uses the variance of the samples to improve the approximation:
\[
\log \hat{p}_{\text{MPS}}(D) \approx \sum_{i=1}^K (\beta_i - \beta_{i-1})\left(\frac{E_{i-1} + E_i}{2}\right)
 - \sum_{i=1}^K \frac{(\beta_i - \beta_{i-1})^2}{12} \left(V_{i} - V_{i+1}\right),
\]
where $V_i = V_{\boldsymbol{\theta} \mid \beta_i}(\log p(D \mid \boldsymbol{\theta}))$ is the variance of the log deviance at $\beta_i$.

\subsection{Stepping stone}
\citet{xie2010improving} proposed the stepping stone (\SStone) algorithm that is related to the path sampling approach described in the previous section.
It uses a series of distributions defining a path between the prior and posterior and therefore inherits the computational burden of path sampling.
Thermodynamic integration and stepping stone differ in the choice of $\beta$ values: \citet{xie2010improving} set $\beta_1, \dots, \beta_n$ equal to the quantiles of a density with fixed parameters (e.g. beta distribution). This approach allows for a more intensive sampling of power posteriors with small $\beta$ values, for which the posterior is changing rapidly.

Let's define the unnormalized power posterior distribution $q_\beta = p(D \mid \boldsymbol\theta)^\beta p(\boldsymbol\theta)$ and normalized power posterior distribution $p_\beta = \frac{q_\beta}{c_\beta}$, where $c_\beta$ is the power marginal likelihood of the data.
The aim of the method is to estimate the ratio $r_{\text{SS}} = c_{1.0}/c_{0.0}$, which is equal to $c_{1.0}$ if the prior is proper.
This ratio  can be expanded into a series of telescopic product of ratios using intermediate power posteriors

\[
r_{\text{SS}} = \frac{c_{1.0}}{c_{0.0}} = \prod_{k=1}^K \frac{c_{\beta_k}}{c_{\beta_{k-1}}} = \prod_{k=1}^K r_{\text{SS},k},
\]
where $r_{\text{SS},k} = c_{\beta_k}/c_{\beta_{k-1}}$ for $k = 1, \dots, K$.
\citet{xie2010improving} estimate each ratio $c_{\beta_k}/c_{\beta_{k-1}}$ by importance sampling using $p_{\beta_{k-1}}$ as the importance distribution.
Using the definition of importance sampling the $k^{th}$ ratio is
\[
\hat{r}_{\text{SS},k} = \frac{1}{n} \sum_{i=1}^n \frac{p(D \mid \boldsymbol{\theta}_{k-1,i})^{\beta_k}}{p(D \mid \boldsymbol{\theta}_{k-1,i})^{\beta_{k-1}}} = \frac{1}{n} \sum_{i=1}^n p(D \mid \boldsymbol{\theta}_{k-1,i})^{\beta_k-\beta_{k-1}},
\]
where $p(D \mid \theta_{k-1,i})$ is the likelihood function evaluated at $\theta_{k-1,i}$, the $i^{th}$ MCMC sample  sampled from $p_{\beta_{k-1}}$.
The product of the $K$ ratios $\hat{r}_{\text{SS},k}$ yields the estimate of the marginal likelihood
\[
\hat{p}_{\text{SS}} = \prod_{k=1}^K \hat{r}_{\text{SS},k}.
\]

\subsection{Generalized stepping stone}
Although stepping stone proved to be more accurate than other approaches,  such as path sampling \citep{xie2010improving}, sampling distributions close to the prior (i.e., small $\beta$ values) can be difficult, particularly if the prior is diffuse.
\citet{fan2010choosing} proposed to generalize the stepping stone method using a reference distribution that approximates the posterior distribution of interest using samples from the posterior distribution to parametrize the reference distribution.
The reference distribution can be independent probability densities from the same family as the prior distribution or the product of densities with the same support.
In our study the priors are exponential distributions, but we used gamma distributions that are parametrized using the method of moments.
The shape and rate parameters are estimated by matching the first two moments of the gamma distribution to the marginal posterior sample mean and variance.

In the same vein as the \SStone\ method, the unnormalized and normalized power posterior distributions in the generalized stepping stone (\GSS) approach are
\begin{align*}
q_\beta &= \big(p(D \mid \boldsymbol\theta) p(\boldsymbol\theta)\big)^\beta \big(p_0(\boldsymbol\theta; \boldsymbol\phi)\big)^{1-\beta},\\
p_\beta &= \frac{q_\beta}{c_\beta},
\end{align*}
where $p(D \mid \boldsymbol\theta)$ is the likelihood function, $p(\boldsymbol\theta)$ is the prior distribution, $p_0$ is the reference distribution parametrized by $\boldsymbol\phi$, and $c_\beta$ is the (power) marginal likelihood of the data.
The key difference with the \SStone\ approach is that for $\beta=0$ the power posterior is equivalent to the reference distribution.

As for the \SStone\ method, the aim of this method is to estimate the ratio $r_{\text{GSS}} = c_{1.0}/c_{0.0}$ using importance sampling.
The ratio $\hat{r}_{\text{GSS}, k}$ is estimated using $n$ samples from $p_{\beta_{k-1}}$:

\[
\hat{r}_{\text{GSS},k} = \frac{1}{n} \sum_{i=1}^n \left( \frac{p(D \mid \boldsymbol\theta_{k-1, i}) p(\boldsymbol\theta_{k-1,i})}{p_0(\boldsymbol\theta_{k-1,i}; \boldsymbol\phi)} \right)^{\beta_k - \beta_{k-1}}.
\]
Combining $\hat{r}_{\text{GSS},k}$ for all $K$ ratios yields the marginal likelihood estimator:
\[
\hat{p}_{\text{GSS}} = \prod_{k=1}^K \hat{r}_{\text{GSS},k}.
\]

\subsection{Nested sampling}
Nested sampling is a Monte Carlo method that aims at calculating the marginal likelihood using a change of variable \citep{skilling2004nested,skilling2006nested}.
It transforms the multidimensional evidence integral over the parameter space into a more manageable one-dimensional integral over the likelihood space.
Skilling defines the prior volume as $dX = p(\boldsymbol{\theta}) d \boldsymbol{\theta}$ so that
\begin{equation}
X(\lambda) = \int _{\mathcal{L}(\boldsymbol{\theta}) > \lambda} p(\boldsymbol{\theta}) d \boldsymbol{\theta},
\label{eq:nsX}
\end{equation}
where $\mathcal{L}(\boldsymbol{\theta})$ is the likelihood function and the integral is taken over the region bounded by the iso-likelihood contour $\mathcal{L}(\boldsymbol{\theta}) = \lambda$.
The marginal likelihood becomes a one-dimensional integral over unit range

\[
p_{\text{NS}}(D) = \int_0^1 L(X) dX,
\]
where $L(X)$ is the inverse function of $X(\lambda)$.

Assuming that $L(X)$ can be computed for a sequence of decreasing values $0 < X_m < \dots < X_0 = 1$, the unit integral can be approximated using quadrature techniques as the weighted sum:
\[
\hat{p}_{\text{NS}}(D) \approx \sum_{i=1}^m L(X_i) w_i,
\]
where $w_i = X_{i} - X_{i-1}$.

The nested sampling algorithm uses a clever process of sampling from the prior (hence $dX$) and conditioning on the likelihood being above a given size (to achieve the likelihood condition of \eqref{eq:nsX}) to approximate the input to such a quadrature technique \citep{skilling2006nested,maturana2018nested}.
The algorithm is initialized with $N$ samples $\{\boldsymbol{\theta}_1, \dots, \boldsymbol{\theta}_N\}$ drawn from the prior and their corresponding likelihoods are calculated $\{\mathcal{L}(\boldsymbol{\theta}_1), \dots, \mathcal{L}(\boldsymbol{\theta}_N)\}$.
The sample with the lowest likelihood $L_{\min}$ is discarded from the set and replaced by a new sample $\boldsymbol{\theta}^*$ drawn from the prior subject to the constraint $L > L_{\min}$.
When we use the discarded point as an $X_i$, the other points in the set of course satisfy the likelihood constraint.
There are a variety of choices for terminating the algorithm \citep{maturana2018nested}.
We choose to terminate when the absolute change in $\log(\hat{p}_{\text{NS}}(D))$ is less than $10^{-6}$.

\subsection{Posterior predictive model selection}
As an alternative to the marginal likelihood, the fit of a model can be assessed through the accuracy of its predictions \citep{gelman1996posterior}.
The probability distribution of a new data set $\tilde{D}$ having observed data set $D$ is defined as
\[
p(\tilde{D} \mid D) = \int p(\tilde{D} \mid \boldsymbol{\theta}) p(\boldsymbol{\theta} \mid D) d \boldsymbol{\theta}.
\]

\subsubsection{Log pointwise predictive density}
A related quantity is the expected log pointwise predictive density \citep{vehtari2017practical} for a new data set, with $n$ data points, is defined as
\[
\text{elpd} = \sum_{i=1}^n \int p_t(\tilde{D}_i) \log p(\tilde{D}_i \mid D) d \tilde{D_i},
\]
where $p_t(\tilde{D}_i)$ is the distribution representing the true data-generating process for $\tilde{D}_i$.
In the phylogenetic framework, the observation $D_i$ corresponds to a single site in the alignment.
Since the $p_t$ is not known, one can use cross-validation to approximate elpd (see next section).

As in \citep{vehtari2017practical}, we define the log pointwise predictive density
\[
\text{lpd} = \sum_{i=1}^n \log p(D_i \mid D) = \sum_{i=1}^n \log \int p(D_i \mid \boldsymbol{\theta}) p(\boldsymbol{\theta} \mid D) d \boldsymbol{\theta},
\]
where $p(D_i \mid \boldsymbol{\theta})$ is the likelihood of the $i^{th}$ observation.
The log pointwise predictive density can be estimated using $S$ draws $\boldsymbol{\theta}_1, \dots, \boldsymbol{\theta}_S$ from the posterior distribution $p(\boldsymbol{\theta} \mid D)$, by summing over the $n$ data points

\[
\widehat{\text{lpd}} = \sum_i^n \log \Big(\frac{1}{S}\sum_{s=1}^S p(D_i \mid \boldsymbol{\theta}_s)\Big), \boldsymbol{\theta}_s \sim p(\boldsymbol{\theta} \mid D).
\]

We compared the fit of our topology models using the predictive accuracy approximation $\widehat{\text{lpd}}$
\[
\log \hat{p}_{\text{PPD}}(D) = \widehat{\text{lpd}}
\]
as an estimate of the log marginal likelihood.
Although we are not aware of others using it in this way, we have found that it provides a reasonable approximation.
However, the lpd of observed data $D$ is an overestimate of the elpd for future data \citep{vehtari2017practical}.

\subsubsection{Conditional predictive ordinates}
A related approach is the conditional predictive ordinates (CPO) method based on Bayesian leave-one-out (LOO).

The leave-one-out estimate of the predictive density for a datapoint is
\[
\text{elpd}_{\text{loo}} = \sum_{i=1}^n \log p(D_i \mid D_{-i}) = \sum_{i=1}^n \log \ \int p(D_i \mid D_{-i}, \boldsymbol{\theta}) p(\boldsymbol{\theta} \mid D_{-i}) d \boldsymbol{\theta},
\]
where $p(D_i \mid D_{-i})$ is the leave-one-out predictive density (aka conditional predictive ordinate) given the data without the $i^{th}$ data point.\\

The CPO estimate of this is given by
\[
\hat{p}(D_i \mid D_{-i}) = \frac{1}{\frac{1}{S}\sum_{i=1}^S \frac{1}{p(D_i \mid \boldsymbol{\theta}_s)}}, \boldsymbol{\theta}_s \sim p(\boldsymbol{\theta} \mid D).
\]
The resulting estimate of the log marginal likelihood (called the log pseudo-marginal likelihood by \citet{lewis2014posterior}) is given by
\[
\log \hat{p}_{\text{CPO}}(D) = \widehat{\text{lpd}_\text{loo}} = \sum_{i=1}^n \log \hat{p}(D_i \mid D_{-i})
\]

\subsection{Variational inference}
Variational Bayes methods provide an analytical approximation to the posterior probability and a lower bound for the marginal likelihood.
The main idea is to choose a family of distributions $q$ parametrised with parameters $\boldsymbol{\phi}$ and to minimize the Kullback Leibler (KL) divergence from variational distribution $q$ to the posterior distribution $p$ of interest

$$ \boldsymbol{\phi}^{*} = \argmin_{\boldsymbol{\phi} \in \boldsymbol{\Phi}}  \mathrm{KL}(q(\boldsymbol{\theta}; \boldsymbol{\phi}) \parallel p(\boldsymbol{\theta} \mid D)). $$

It is difficult to minimise the KL divergence directly but much easier to minimize a function that is equal to it up to a constant.
Expanding the KL divergence we get

$$\begin{aligned}
\mathrm{KL}(q(\boldsymbol{\theta}; \boldsymbol{\phi}) \parallel p(\boldsymbol{\theta} \mid D)) &= \mathop{\mathbb{E}}[\log q(\boldsymbol{\theta}; \boldsymbol{\phi})] - \mathop{\mathbb{E}}[\log p(\boldsymbol{\theta} \mid D)] \\
  & = \mathop{\mathbb{E}}[\log q(\boldsymbol{\theta}; \boldsymbol{\phi})] - \mathop{\mathbb{E}}[\log p(\boldsymbol{\theta}, D)] + \log p(D)\\
    & = -\textrm{ELBO}(\boldsymbol{\phi}) + \log p(D),
\end{aligned}$$
where $\textrm{ELBO}(\boldsymbol{\phi}) = \mathop{\mathbb{E}}[\log p(\boldsymbol{\theta}, D)] - \mathop{\mathbb{E}}[\log q(\boldsymbol{\theta}; \boldsymbol{\phi})]$.
This equation suggests that the $\textrm{ELBO}(\boldsymbol{\phi})$ is the lower bound of the evidence: $\log p(D) \geq \textrm{ELBO}(\boldsymbol{\phi})$.

Instead of minimizing KL divergence, we maximize  the evidence lower bound:
$$ \textrm{ELBO}(\boldsymbol{\phi}) = \mathop{\mathbb{E}}_{q(\boldsymbol{\theta}; \boldsymbol{\phi})}[\log p(D, \boldsymbol{\theta}) - \log q(\boldsymbol{\theta}; \boldsymbol{\phi})].$$

Several variational distributions can be used including the mean-field and fullrank Gaussian distributions.
The fullrank model uses a multivariate Gaussian distribution to model the correlation between variables while the meanfield distribution assumes a diagonal covariance matrix.
In this study we used the meanfield model hence taking the assumption that there is no correlation between the branch lengths of the phylogeny:
\[
q(\boldsymbol{\theta}; \boldsymbol{\phi}) = \mathcal{N}(\boldsymbol{\theta}; \boldsymbol{\mu}, \diag(\boldsymbol{\sigma}^2)) = \prod_{i=1}^n \mathcal{N}(\theta_i; \mu_i, \sigma_i^2).
\]

It is common to use stochastic gradient ascent algorithm to maximise the ELBO as long as the model is differentiable  \citep{ranganath2014black,kucukelbir2015automatic}.
In the phylogenetic context the derivative of posterior with respect to the branch lengths can be derived analytically without resorting to approximations such as finite differences.
We used a log transform on the branch lengths to ensure that the variational distribution stays within the support of the posterior.

Given an optimized variational model we used the ELBO as an approximation of the marginal likelihood
\[
\hat{p}_{\text{ELBO}}(D)  = \max_{\boldsymbol{\phi} \in \boldsymbol{\Phi}}\textrm{ELBO}(\boldsymbol{\phi}).
\]
%where $\mathfrak{Q}$ is a family of distributions parameterized by $\boldsymbol{\phi}$ over the latent variables $\boldsymbol{\theta}$.

The ELBO estimates can have high variance and might be of little use to discriminate between closely related models (in the KL sense).
We used importance sampling to calculate the marginal likelihood of a model using the variational distribution $q$ as the importance distribution. This yields the $\hat{p}_{\text{VBIS}}(D)$ estimator:

\[
\hat{p}_{\text{VBIS}}(D) = \frac{1}{N} \sum_{i=1}^N \frac{p(D \mid \tilde{\boldsymbol{\theta}_i}) p(\tilde{\boldsymbol{\theta}_i})}{q_{\text{ELBO}}(\tilde{\boldsymbol{\theta}_i})}, \tilde{\boldsymbol{\theta}}_i \sim q_{\text{ELBO}}(\boldsymbol{\theta}).
\]

\section{Supplementary Figures}

For completion, we include here equivalents of Figure \ref{fig:approx_vs_srf} and Figure \ref{fig:split_probs} for datasets DS1-4.
We also include versions of Figure \ref{fig:time_vs_accuracy} and Figure \ref{fig:rmsd_by_ds} that use KL divergence instead of RMSD as the measure of accuracy.
The KL and RMSD results are qualitatively similar.

\begin{figure}[H]
\begin{center}
\includegraphics[width=0.8\linewidth]{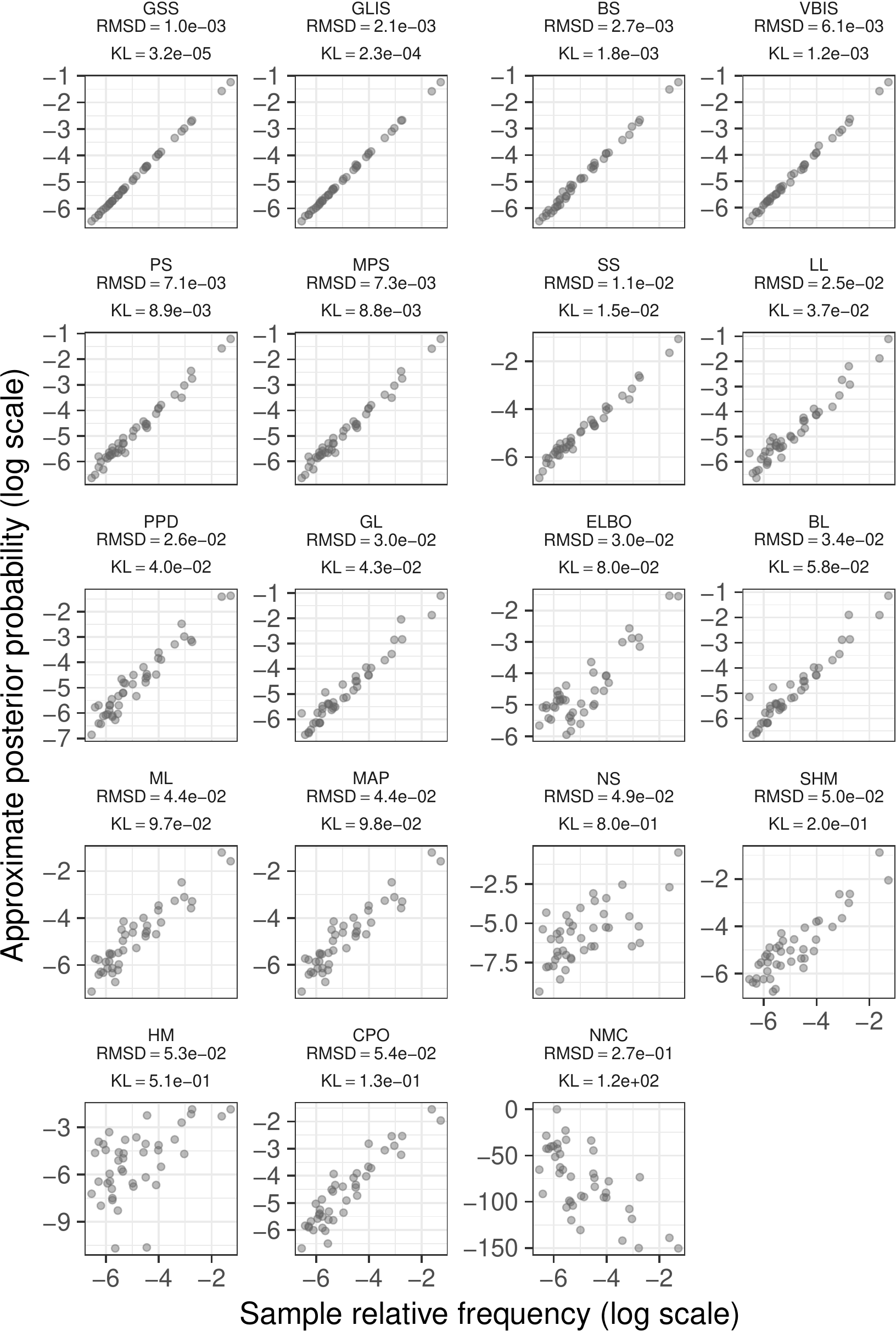}
\end{center}
\caption{The approximate posterior probabilities of the topologies in DS1 versus the ground truth posterior probabilities from MrBayes, plotted on the log scale for clarity.
Results are for a single run of each method.
Panels are ordered by RMSD in increasing order.}
\label{fig:approx_vs_srf_1}
\end{figure}

\begin{figure}[H]
\begin{center}
\includegraphics[width=0.8\linewidth]{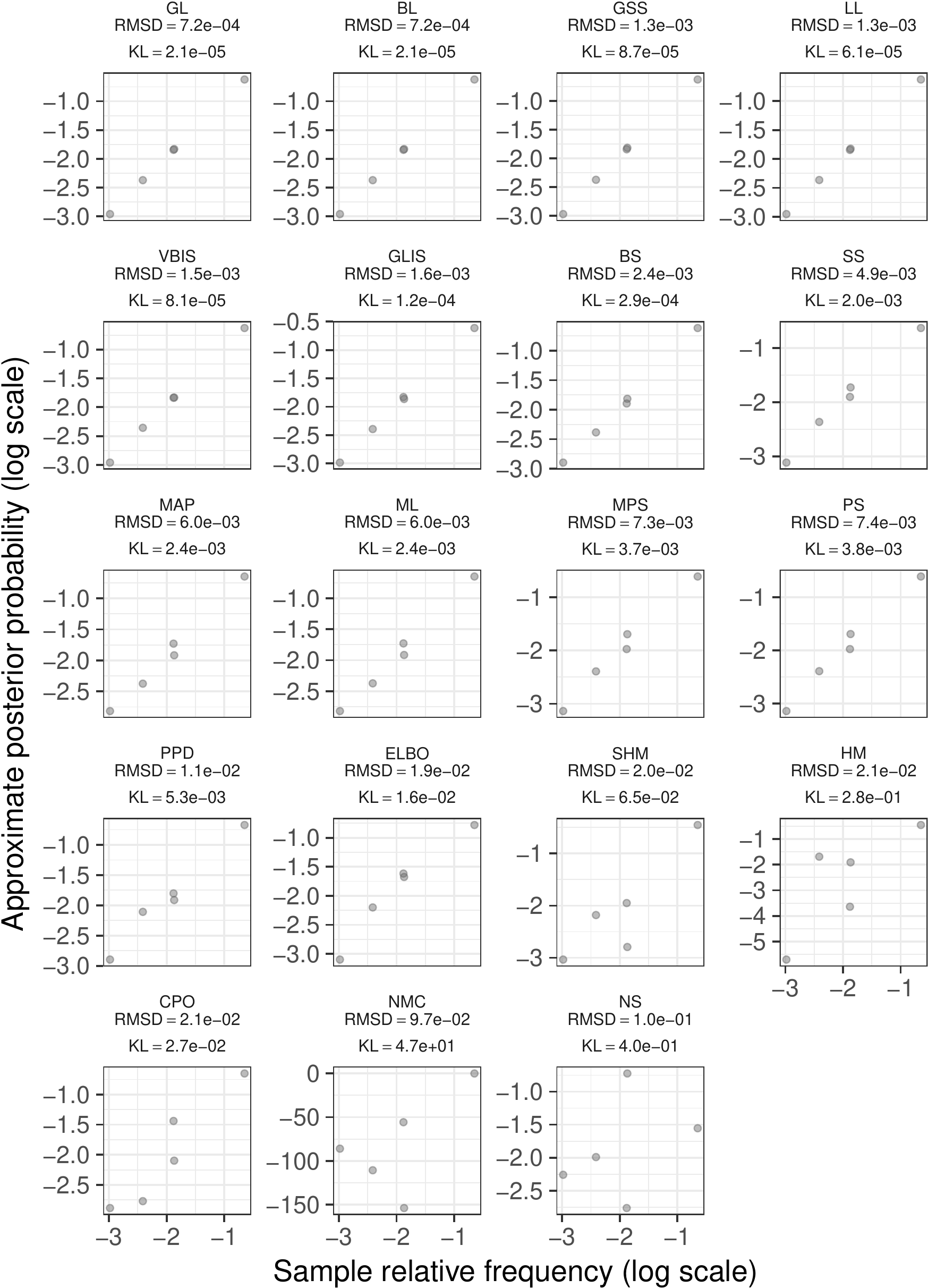}
\end{center}
\caption{The approximate posterior probabilities of the topologies in DS2 versus the ground truth posterior probabilities from MrBayes, plotted on the log scale for clarity.
Results are for a single run of each method.
Panels are ordered by RMSD in increasing order.}
\label{fig:approx_vs_srf_2}
\end{figure}

\begin{figure}[H]
\begin{center}
\includegraphics[width=0.8\linewidth]{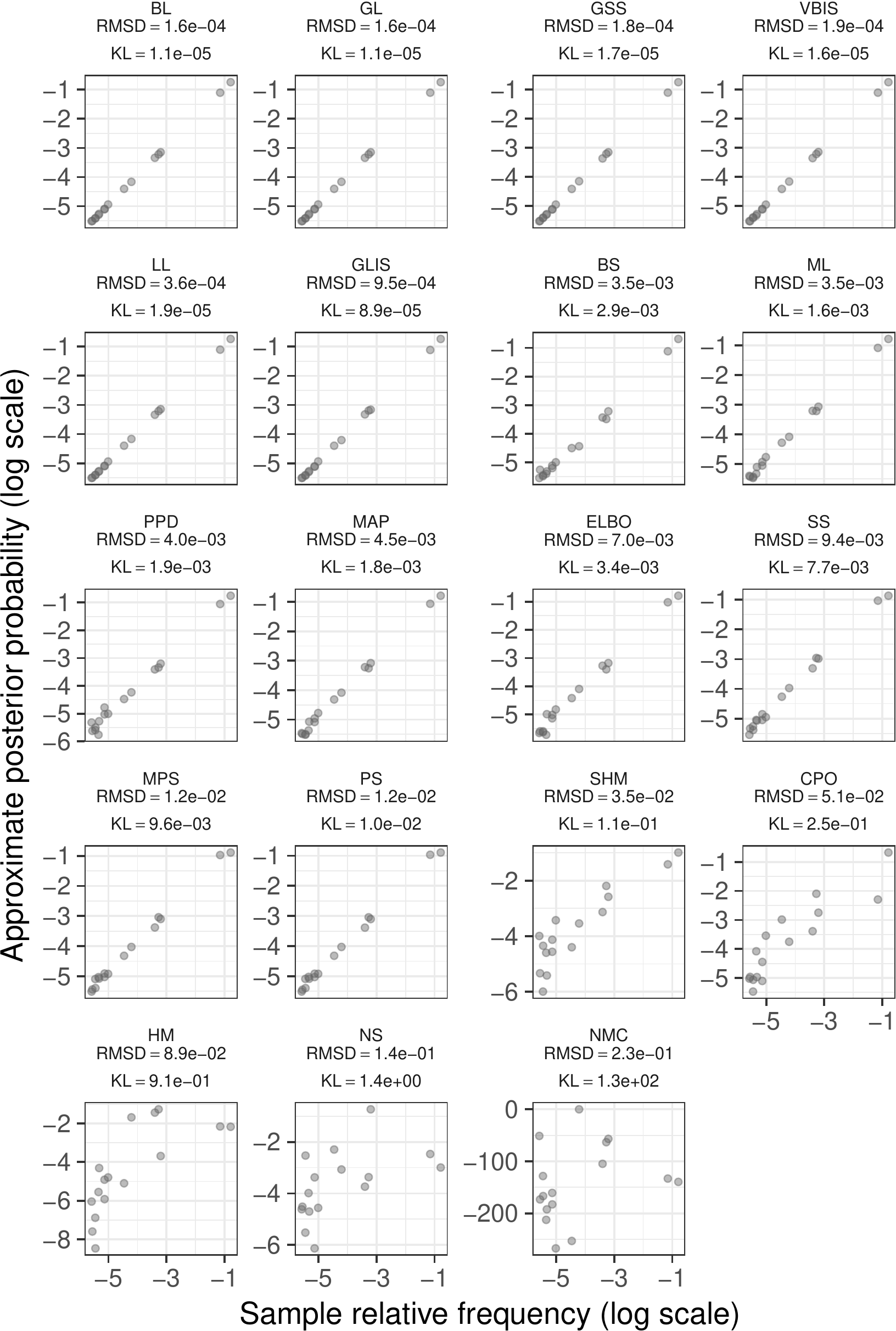}
\end{center}
\caption{The approximate posterior probabilities of the topologies in DS3 versus the ground truth posterior probabilities from MrBayes, plotted on the log scale for clarity.
Results are for a single run of each method.
Panels are ordered by RMSD in increasing order.}
\label{fig:approx_vs_srf_3}
\end{figure}

\begin{figure}[H]
\begin{center}
\includegraphics[width=0.8\linewidth]{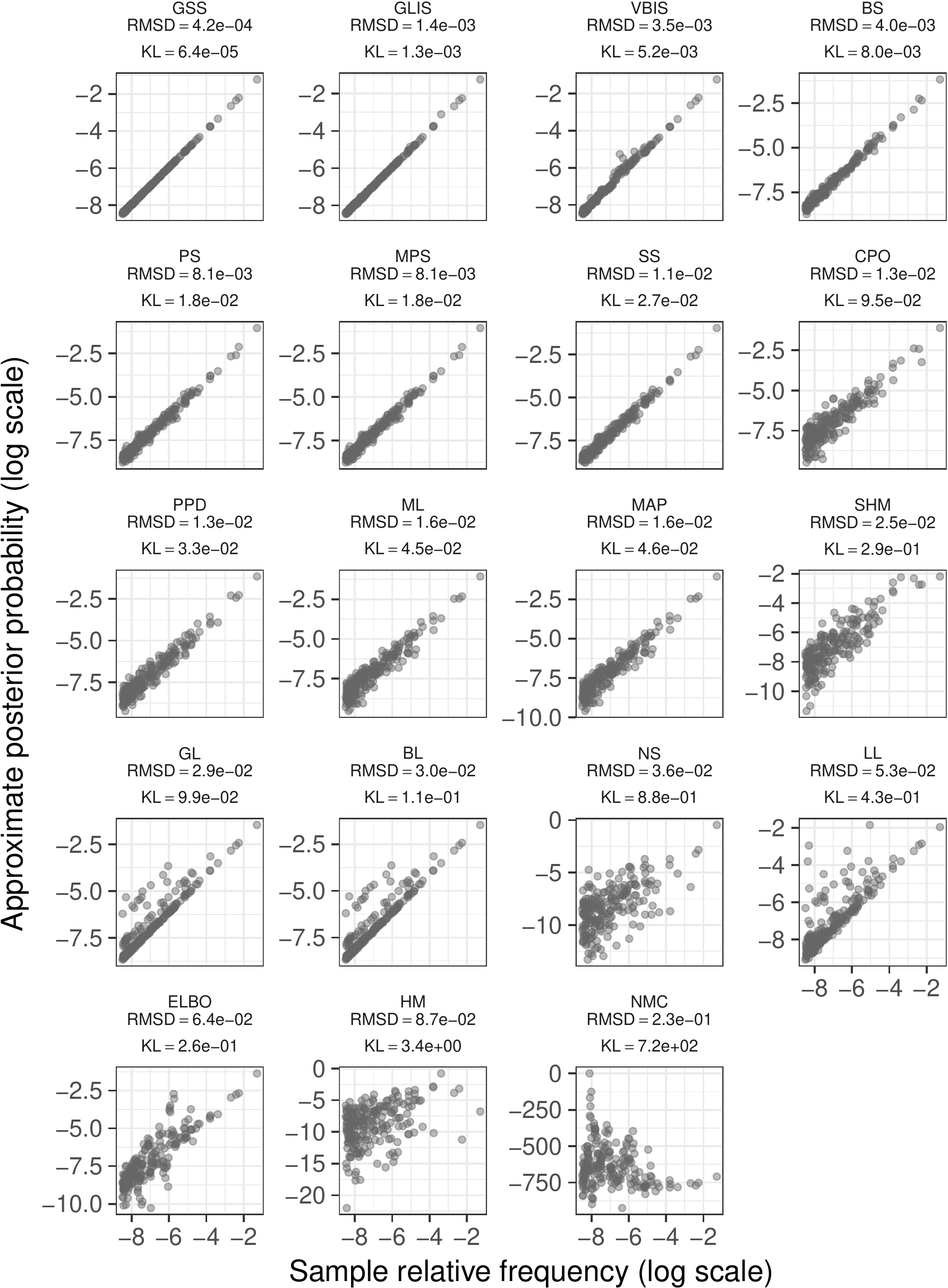}
\end{center}
\caption{The approximate posterior probabilities of the topologies in DS4 versus the ground truth posterior probabilities from MrBayes, plotted on the log scale for clarity.
Results are for a single run of each method.
Panels are ordered by RMSD in increasing order.}
\label{fig:approx_vs_srf_4}
\end{figure}

\begin{figure}[H]
\begin{center}
\includegraphics[width=0.8\linewidth]{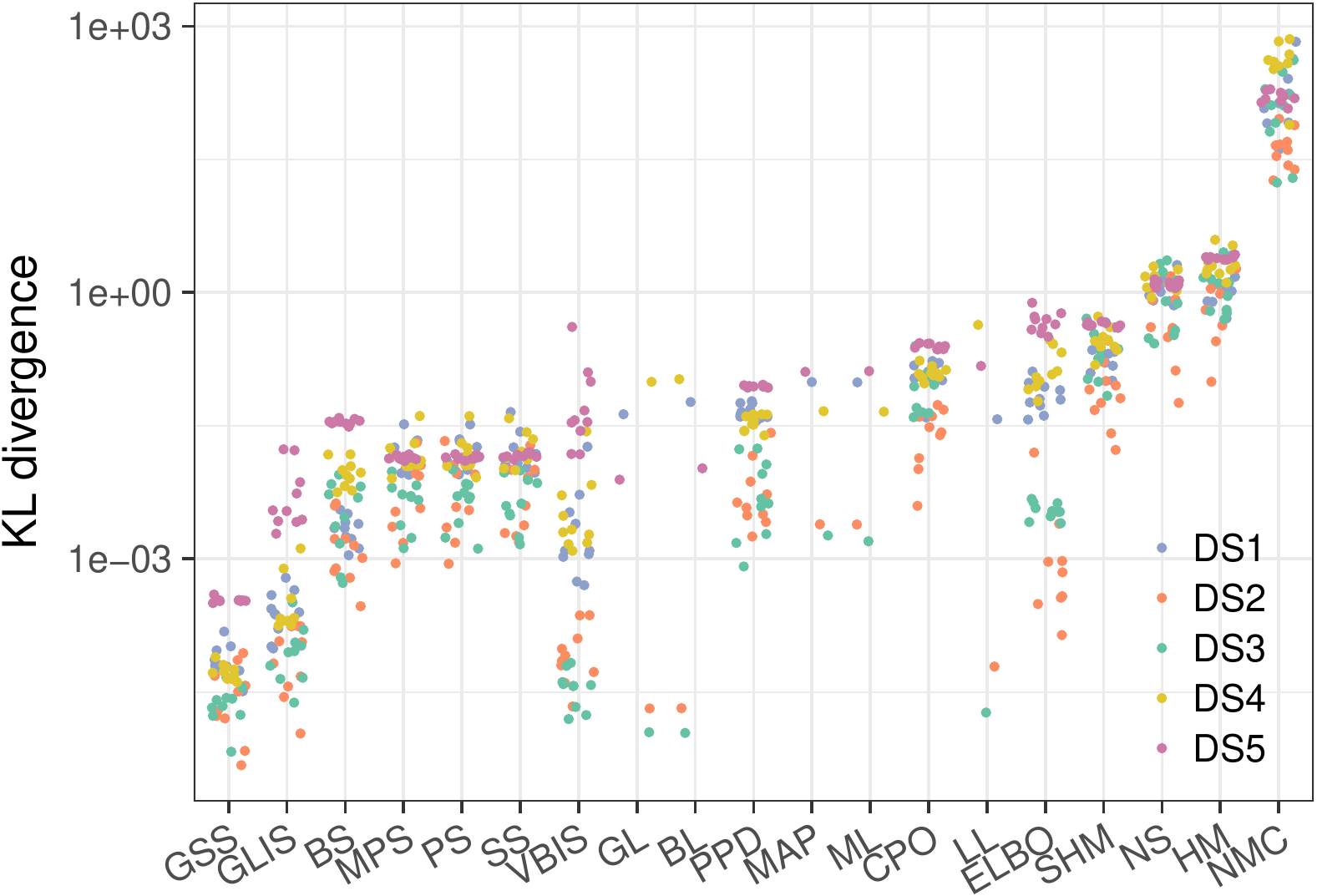}
\end{center}
\caption{Average Kullback-Leibler (KL) divergence from MrBayes posteriors to approximate posteriors for each method on each dataset for 10 replicates.  \LL, \GL,  BL, MAP, and ML are deterministic and therefore only one replicate is shown.}
\label{fig:kl_divergences}
\end{figure}

\begin{figure}[H]
\begin{center}
\includegraphics[width=0.8\linewidth]{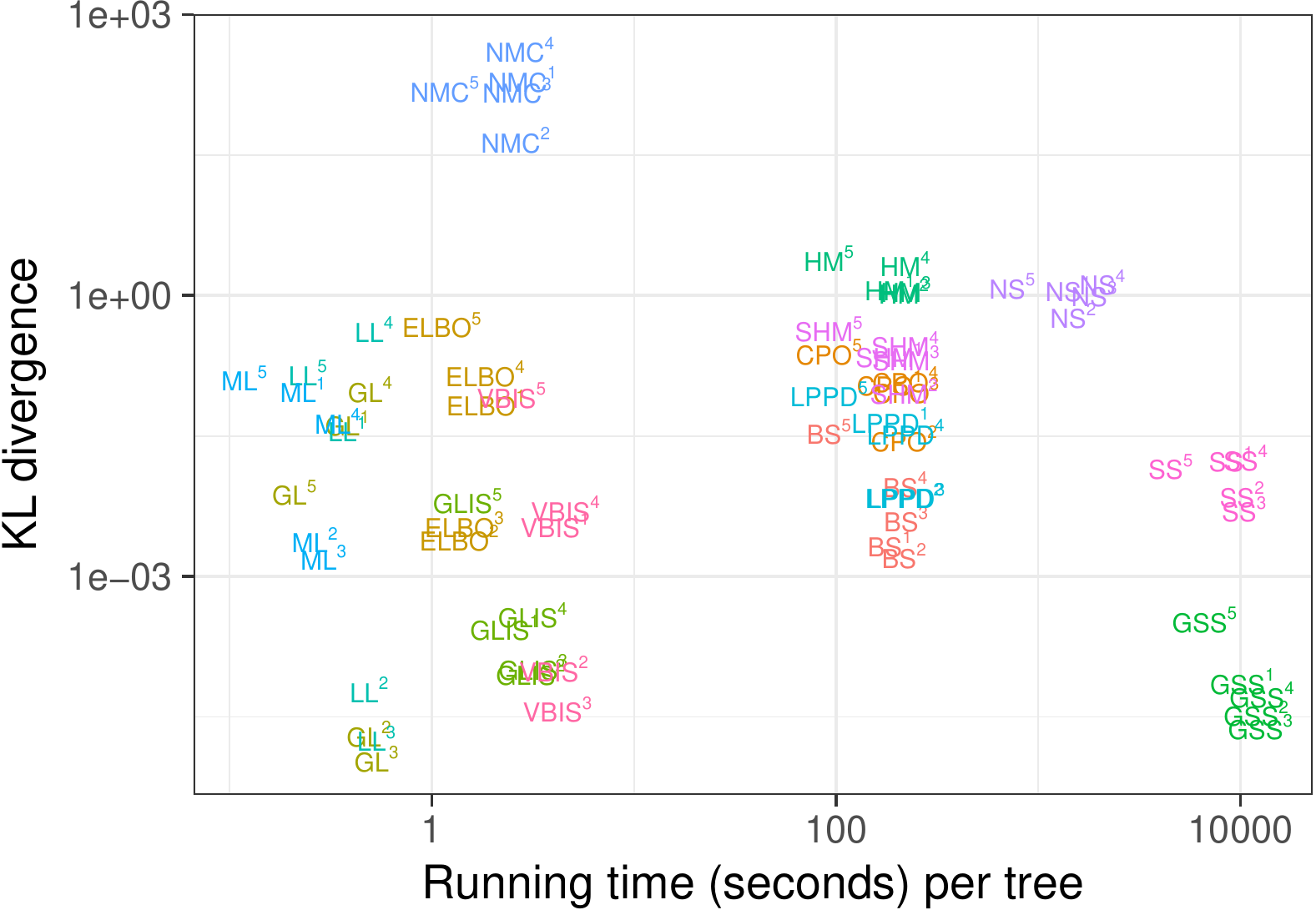}
\end{center}
\caption{Average Kullback-Leibler (KL) divergence from MrBayes posteriors to approximate posteriors of splits in the approximate posterior against running  time. Text denotes method used, while superscripts label applications to individual datasets. Four methods are omitted for visual clarity: MAP is essentially identical to ML, BL is nearly identical to \GL, and \PS\ and \PStwo\ are both similar to \SStone. The horizontal dashed and solid lines depict RMSDs of 0.01 and 0.05 respectively. The KL divergence is calculated using the average marginal likelihood of each tree from each of 10 replicate analyses. The running time is calculated using the average running time of each tree from each of 10 replicate analyses.}
\label{fig:time_vs_accuracy_kl}
\end{figure}

%\bibliographystyle{unsrtnat}
%\bibliography{main}

\end{document}